\documentclass[journal]{IEEEtran}
\usepackage{enumerate}
\usepackage{cite}
\usepackage{graphicx}
\usepackage{alltt}
\usepackage{amsmath}
\usepackage{amsopn}
\usepackage{amssymb}
\usepackage[table]{xcolor}

\newtheorem{lemma}{Lemma}
\newtheorem{proposition}{Proposition}
\newtheorem{corollary}{Corollary}

\newtheorem{definition}{Definition}
\newtheorem{claim}{Claim}

\DeclareMathOperator{\mincut}{mincut}

\DeclareMathOperator{\out}{out}
\DeclareMathOperator{\inp}{in}

\DeclareMathOperator{\DC}{DC}
\DeclareMathOperator{\mbr}{MBR}

\DeclareMathOperator{\GF}{GF}

\newcommand{\gf}{\mathcal{G}_F}

\DeclareMathOperator{\rank}{rank}
\DeclareMathOperator{\sspan}{span}

\title{Locally Repairable Regenerating Codes: Node Unavailability and the Insufficiency of Stationary Local Repair}

\author{Imad~Ahmad,~\IEEEmembership{Student Member,~IEEE,}
and~Chih-Chun~Wang,~\IEEEmembership{Senior Member,~IEEE}
\thanks{This work was supported in parts by NSF grants CCF-0845968, CNS-0905331, CCF-1422997, and ECCS-1407604. Part of the results was presented in the 2015 IEEE International Symposium on Information Theory.}

\thanks{I. Ahmad and C.-C. Wang are with the School of Electrical and Computer Engineering, Purdue University, West Lafayette,
IN, 47906 USA e-mail: \{ahmadi,chihw\}@purdue.edu.}

}
\begin{document}
\maketitle
\thispagestyle{plain}
\pagestyle{plain}

\begin{abstract}
Locally repairable codes (LRCs) are ingeniously designed distributed storage codes with a (usually small) fixed set of helper nodes participating in repair. Since most existing LRCs assume {\em exact repair} and allow {\em full exchange of the stored data} ($\beta=\alpha$) from the helper nodes, they can be viewed as a generalization of the traditional erasure codes (ECs) with a much desired feature of local repairability via predetermined sets of helpers. However, it also means that they lack the features of (i) functional repair, and (ii) partial information-exchange ($\beta<\alpha$) in the original regenerating codes (RCs), which could further reduce the repair bandwidth. Motivated by the significant bandwidth reduction of RCs over ECs, existing works by Ahmad \emph{et al} and by Hollmann studied the concept of ``locally repairable regenerating codes (LRRCs)'' that successfully combine functional repair and partial information exchange of regenerating codes with the much-desired local repairability feature of LRC. The resulting LRRCs demonstrate significant bandwidth reduction.

One important issue that needs to be addressed by any local repair schemes (including both LRCs and LRRCs) is that sometimes designated helper nodes may be temporarily unavailable, the result of multiple failures, degraded reads, or other network dynamics. Under the setting of LRRCs with temporary node unavailability, this work studies the impact of different helper selection methods. It proves that with node unavailability, all existing methods of helper selection, including those used in RCs and LRCs, can be insufficient in terms of achieving the optimal repair-bandwidth. For some scenarios, it is necessary to combine LRRCs with a new helper selection method, termed {\em dynamic helper selection}, to achieve optimal repair bandwidth. This work also compares the performance of different helper selection methods and answers the following fundamental question: \emph{whether one method of helper selection is intrinsically better than the other?} for various different scenarios.
\end{abstract}

\section{Introduction}

\begin{table*}[t]
\caption{The comparison table between existing codes and LRRCs of this work.}
\begin{center}
\begin{tabular}{| p{2.5cm} || p{2cm} | p{2cm} | p{2cm} |p{3.7cm}|p{2cm}|}
\hline
    &Repair Mode & Info-Exchange& Helper Selection& Temporary Node Unavailability&Target Parameters\\
\hline\hline

ECs& Exact & Full & Blind& No Treatment Needed& $d=k$\\ \hline
Original RCs& \cellcolor{blue!25}Functional & \cellcolor{blue!25}Partial & Blind& No Treatment Needed& $d\geq k$\\ \hline
Exact Repair RCs& Exact & \cellcolor{blue!25}Partial & Blind& No Treatment Needed& $d\geq k$\\ \hline
LRCs& Exact & Full & \cellcolor{blue!25}Intelligent & Need Special Treatment& $d<k$\\ \hline
LRRCs of this work& \cellcolor{blue!25}Functional & \cellcolor{blue!25}Partial & \cellcolor{blue!25}Intelligent & Need Special Treatment& All Parameters\\ \hline

\end{tabular}
\label{table1}
\end{center}
\end{table*}

\par Erasure coding (EC) is efficient in terms of reliability versus redundancy tradeoff in distributed storage systems. An $(n,k)$ MDS code, when applied over a network of $n$ storage nodes, can tolerate $n-k$ simultaneous failures. When a node fails, it is repaired by accessing any $k$ {\em surviving nodes}, downloading all the coded data stored in these $k$ nodes, and then reconstructing the original data. As a result, we say that the repair of EC involves ``full information exchange'' and ``exact repair''. 

\par Regenerating codes (RCs) \cite{dimakis2010network}, on the other hand, were proposed to decrease the amount of communication required during repair, oftentimes termed the \emph{repair-bandwidth}. The three key ideas that allow regenerating codes to decrease repair-bandwidth are: (i) contact as many nodes as possible during repair or in other words $d=n-1$ nodes, termed \emph{helper nodes}, (ii) download only a partial fraction of the data ($\beta<\alpha$) as opposed to full information-exchange ($\beta=\alpha$), and (iii) allow {\em functional repair} which is a generalization to exact repair in ECs. These 3 ideas enable significant repair-bandwidth reduction of RCs over ECs \cite{dimakis2010network}. 

\par Another type of distributed storage codes is the locally repairable codes (LRCs) \cite{gopalan2012locality,prakash2012optimal,papailiopoulos2014locally,kamath2014codes,rawat2014optimal,kamath2013explicit} that use a small number of helper nodes $d$ during repair, which is in contrast with RCs that were originally designed for large $d$ (i.e., $d\geq k$). A closer look at the properties of LRCs shows that LRCs resemble ECs in that they operate with $\alpha=\beta$ and under exact repair. The main difference\footnote{Another subtle difference is that for ECs, a newcomer can access {\em any} set of $d$ helpers while for LRCs each newcomer can only access {\em a predetermined set} of $d$ helpers. As will be rigorously defined in Section~\ref{subsec:types_schemes}, the former is termed the {\em blind helper selection} and the latter is called the {\em stationary helper selection}.} is that ECs access $d=k$ helpers while LRCs use much smaller $d$ values (usually $\ll k$). For that reason, LRCs can be viewed as a generalization of ECs with a much desired feature of local repairability (small $d$). Inspired by the repair-bandwidth reduction of RCs over ECs, it is thus natural to ask the question: whether it is possible to design {\em locally repairable regenerating codes} (LRRCs), that simultaneously admit all three features: local repairability ($d<k$), partial information-exchange ($\beta <\alpha$), and functional repair?  

\par It is worth noting that for any locally repairable code (small $d$) how to select the $d$ helpers (out of the remaining $n-1$ nodes) is a critical part of the underlying code design and could have significant impact on its performance. Therefore, any attempt on designing LRRCs that combine local repairability ($d<k$), partial information-exchange, and functional repair must address the challenges of how to design the underlying helper selection policy.

\par It turns out that the original work \cite{dimakis2010network} that proposed RCs does consider the possibility of LRRCs since it derives the storage-repair-bandwidth tradeoff curves for arbitrary $d<n-1$. However, a close look at the derivation in \cite{dimakis2010network} shows that \cite{dimakis2010network} assumes that the LRRC ``blindly chooses the $d$ helpers'' and then characterizes the corresponding worst-case performance under such a \emph{blind helper selection} (BHS). In some sense, \cite{dimakis2010network} has, implicitly, analyzed LRRCs under BHS, the most pessimistic helper selection scheme. It was not clear from the results in \cite{dimakis2010network} whether the performance of LRRCs can be further improved if some sophisticated helper selection scheme other than BHS is used.

\par In contrast, existing works in \cite{ahmad2014when,arxiv1,arxiv2} and \cite{hollmann2014minimum} are the first works to study LRRCs when intelligent (non-blind) helper selection policies are used. For the special cases of $k=n-1$ and $\alpha=d\beta$ or $\alpha=\beta$, \cite{hollmann2014minimum} proves a lower bound on the repair bandwidth (BW) of any LRRCs regardless whether an intelligent or a BHS scheme is used. The lower bound turns out to be tight and achievable by some modified LRC scheme in \cite{papailiopoulos2014locally} for certain $(n,k,d)$ combinations. 

\par At the same time, \cite{ahmad2014when,arxiv1,arxiv2} answer the following related question: under what $(n,k,d)$ values can intelligent helper selection strictly improve the performance of LRRCs when compared to the BHS-based LRRC in \cite{dimakis2010network}. This question was fully answered for any arbitrary $(n,k,d)$ parameters. A new scheme termed the \emph{family helper selection} (FHS) scheme was also devised in \cite{ahmad2014when,arxiv1,arxiv2} that demonstrates superior performance (very small repair BW) while admitting local repairability. The FHS scheme proposed in \cite{ahmad2014when,arxiv1,arxiv2} is provably optimal (i.e., attains the upper bounds in \cite{ahmad2014when,arxiv1,arxiv2} and \cite{hollmann2014minimum}) for a much wider range\footnote{The scheme in \cite{papailiopoulos2014locally} can be viewed as a special example of the FHS scheme and its variant in \cite{ahmad2014when,arxiv1,arxiv2}.} of $(n,k,d)$ values than the modified LRC scheme in \cite{papailiopoulos2014locally}.

Table~\ref{table1} summarizes the differences among ECs, RCs, LRCs, and LRRCs in terms of repair modes, the amount of information exchange, the corresponding helper selection schemes, and the target parameter values. The blue-shaded blocks correspond to the ideas that are known to be able to reduce the repair BW. Note that only LRCs and LRRCs employ intelligent helper selection rules while both ECs and RCs employ blind helper selection. 

\par Despite the preliminary promising results, the LRRCs considered in \cite{ahmad2014when,arxiv1,arxiv2} and \cite{hollmann2014minimum} do not consider the following practical issue: Because of multiple failures or degraded reads or other network dynamics, some designated helper nodes may be temporarily unavailable. Therefore, for any locally repairable scheme to work in practice, including both LRCs and LRRCs, it needs to have an alternative set of helpers in case of node unavailability. See the column titled ``temporary node unavailability'' in Table~\ref{table1}. For LRCs, temporary node unavailability has been studied in \cite{prakash2012optimal,rawat2014optimal,pamies2013locally}. In this work, we study the performance of LRRCs \cite{ahmad2014when,arxiv1,arxiv2} and \cite{hollmann2014minimum} under different helper selection policies while taking into account the issue of temporary node unavailability.

\par Our studies are centered around three different classes of helper selection schemes. (i) the BHS schemes; (ii) the stationary helper selection (SHS) schemes; and (iii) a new class of schemes proposed in this work, called {\em dynamic helper selection} (DHS). These three classes of helper selection schemes will be formally defined in Section~\ref{subsec:types_schemes}. As will be explained in details in Section~\ref{subsec:helper_existing}, {\em the helper selection policies of all existing designs of ECs, RCs, LRCs, and LRRCs \cite{dimakis2010network, gopalan2012locality,prakash2012optimal,papailiopoulos2014locally,kamath2014codes,rawat2014optimal,kamath2013explicit,ahmad2014when,arxiv1,arxiv2,hollmann2014minimum} are either BHS or SHS}.

The main contributions of this work are summarized as follows. 

{\bf Contribution~1:} We prove, for the first time in the literature, that both BHS and SHS can be insufficient in terms of achieving the optimal repair-bandwidth. Specifically, we provide an example with $r=1$ showing that it is necessary to use DHS, which is designed based on a completely different principle, to achieve optimal repair-bandwidth while the performance of BHS and any SHS are strictly suboptimal. Furthermore, the DHS scheme in our example is simultaneously minimum bandwidth regenerating (MBR) and minimum storage regenerating (MSR), attaining a new storage-BW tradeoff point that was previously believed to be not possible except for some trivial degenerate cases. Such an example demonstrates the benefit of DHS and calls for further research participation on DHS designs.

{\bf Contribution~2:} Being a blind scheme, BHS is the least powerful of the three helper selection policies and can thus be used as a baseline. We study the following fundamental question: Given any $(n,k,d,r)$ value, whether we can design an SHS or DHS scheme that strictly outperforms BHS? Surprisingly, for many $(n,k,d,r)$ values the answer is no. That is, for those $(n,k,d,r)$ values, even the best SHS or DHS scheme is no better than the simple BHS solution used in the original RCs \cite{dimakis2010network}. We call those $(n,k,d,r)$ values as being {\em indifferent to helper selection} since the performance does not depend on what type of helper selection schemes being used. 
 
\par Knowing whether a given $(n,k,d,r)$ value is indifferent to helper selection is of significant practical value since a distributed storage code designer can then decide whether to simply use the most basic BHS scheme (if the underlying $(n,k,d,r)$ is indifferent to helper selection) or to invest time and effort to design more sophisticated helper selection rules to further improve the performance of the system. 

\par In this work, we prove that for a vast majority of $(n,k,d,r)$ values, we can answer unambiguously whether it is indifferent to helper selection or not by checking some very simple conditions. 

\par {\bf Summary:} The main contribution of this work is mostly information-theoretic. The carefully constructed example sheds surprisingly new insights on the fundamental performance limits of different helper selection schemes in the context of RCs and LRRCs. The results in Contribution~2 allows us to quickly check whether a given $(n,k,d,r)$ value is indifferent to helper selection or not, which provides valuable case-by-case guidelines whether it is beneficial to spend time designing new SHS or DHS schemes or whether one should simply use the simple BHS. 

The rest of this paper is organized as follows. Section~\ref{sec:problem_statement} motivates the problem and introduces key definitions and notation. Section~\ref{sec:ifg} describes information-flow graphs, the main tool we used for the analysis of LRRCs. Section~\ref{sec:main_results} presents the main results of this work. Section~\ref{sec:dy_better} presents proofs of the results of Contribution~1. Section~\ref{sec:mincut} presents the proofs of the results of Contribution~2. Section~\ref{sec:conclusion} concludes this work.

\section{Problem Statement}\label{sec:problem_statement}

\subsection{The Parameters of A Distributed Storage Network} \label{subsec:param}
\par This work follows the same distributed storage network model as introduced in the seminal work \cite{dimakis2010network}. For completeness, we provide in the following detailed definitions of some key parameters. Further explanations of the system model can be found in \cite{dimakis2010network}.

{\bf Parameters $n$ and $k$:} We denote the total number of nodes in a storage network by $n$. For any $1\leq k\leq n-1$, we say that a code can satisfy the reconstruction requirement if any $k$ nodes can be used to reconstruct the original data/file. For example, consider a network of 7 nodes. A $(7,4)$ Hamming code can be used to protect the data. We say that the Hamming code can satisfy the reconstruction requirement for $k=6$. Since any 6 nodes can construct the original file. By the same definition, the Hamming code can also satisfy the reconstruction requirement for $k=5$ and $k=4$, but cannot satisfy the reconstruction requirement for $k=3$. The smallest $k$ of the $(7,4)$ Hamming code is thus $k^*=4$. In general, the value of $k$ is related to the {\em desired} protection level of the system while the value of $k^*$ is related to {\em actual} protection level offered by the specific distributed storage code implementation.

\par For example, suppose the design requirement is $k=6$, we can still opt for using the $(7,4)$ Hamming code to provide the desired level of protection. However, using $(7,4)$ Hamming code may be an overkill since a $(7,4)$ Hamming code has $k^*=4$ and it is possible to just use a single-parity bit to achieve $k=6$.

\par {\bf Parameter $d$:} We denote the number of nodes that a newcomer can access during repair by $d$. For example, \cite{dimakis2010network} provides a detailed RC construction about how to achieve the design goal $(n,k,d)=(10,7,9)$. Namely, each newcomer can access $d=9$ helpers and any $k=7$ nodes can be used to reconstruct the original file. At the same time, \cite{dimakis2010network} also provides high-level guidelines how to use the RC to achieve the design goal when $(n,k,d)=(10,7,5)$. However, the RC can be an overkill in this scenario $(n,k,d)=(10,7,5)$ since any RC construction in \cite{dimakis2010network} that can achieve $(n,k,d)=(10,7,5)$ can always achieve $k^*= d=5$. As a result, even though the high-level design goal is to only protect against $10-7=3$ failures under the constraint of accessing only $d=5$ helpers during repair, the RC in \cite{dimakis2010network} cannot take advantage of this relatively loose protection-level requirement since it always has $k^*\leq d=5$.

\par Note that the above observation does not mean that the system designer should never use the RCs \cite{dimakis2010network} when the design goal is $(n,k,d)=(10,7,5)$. The reason is that these RCs with BHS have many other advantages that may be very appealing in practice, e.g., some very efficient algebraic code construction methods \cite{shah2012interference}, allowing repair with $(n-d)$ simultaneous failures, and admitting efficient collaborative repair when more than one node fails \cite{shum2013cooperative}. The fact that $k^*\leq d$ for any RCs in \cite{dimakis2010network} simply means that when the requirement is $(n,k,d)=(10,7,5)$, the system designer should be aware that the RCs with BHS in \cite{dimakis2010network} cannot take full advantage of the relatively loose required protection level since we have in this scenario $k>d\geq k^*$.

\par In this work, we focus on the design target $k$ instead of the actual performance parameter $k^*$, since given the same $k$, the actual $k^*$ value may depend on how we implement the codes. For example, when locally repairable codes \cite{gopalan2012locality} are used, it is possible to design a system with $k=k^*>d$. However, when RCs are used together with BHS, we always have $k^*\leq d$ even though the target protection level may satisfy $k>d$. For any given $(n,k,d)$ values, the goal of this paper is to compare the best performance of any possible helper selection scheme that can still satisfy the desired $(n,k,d)$ values regardless whether they offer over-protection ($k> k^*$) or not.

{\bf Parameter $r$:} We denote the maximum number of nodes that can be temporarily unavailable at any given time by $r$. Specifically, if we denote the set of unavailable nodes by $U$, then we must have $|U|\leq r$. If we also denote the failed node by $F$, the design goal is to repair node $F$ when the nodes in $U$ are unavailable. The unavailability of nodes in $U$ may be due to degenerate reads, multiple failures, or underlying network dynamics. In this work we do not consider repair collaboration. That means, even when we have multiple failures, say both nodes $i$ and $j$ fail simultaneously, we repair each node separately. For example, we set $F=i$ and $U=\{j\}$ when repairing node $i$ and we set $F=j$ and $U=\{i\}$ when repairing node $j$. Some repair cooperation schemes that jointly repair both nodes $i$ and $j$ can be found in \cite{shum2013cooperative}.

\par {\bf The range of the design criteria $(n,k,d,r)$:} Due to the nature of the distributed storage problem, we only consider $(n,k,d,r)$ values that satisfy
\begin{align} \label{eq:par_cond}
2\leq n;~ 1\leq k \leq n-1;~ 1\leq d; \text{ and } d\leq n-1-r.
\end{align}
In all the results in this work, we assume {\em implicitly} that the $n$, $k$, and $d$ values satisfy \eqref{eq:par_cond}.

\par {\bf Parameters $\alpha$, $\beta$, and $\mathcal{M}$:} The overall file size is denoted by $\mathcal{M}$. The storage size for each node is $\alpha$, and during the repair process, the newcomer requests $\beta$ amount of traffic from each of the helpers. The total repair-bandwidth is thus $\gamma\stackrel{\Delta}{=}d\beta$.

\subsection{Types of Helper Selection Schemes}\label{subsec:types_schemes}
\par {\bf DHS:} We consider the most general form of helper selection in which the helper selection at current time $\tau$ can depend on the time index $\tau$ and the history of node failures and node unavailability from all the previous time slots 1 to ($\tau-1$). We term this type of schemes the \emph{dynamic helper selection (DHS)} scheme. Mathematically, for every time slot $\tau$ (or equivalently for the $\tau$-th repair) the helper set decision at time $\tau$ can be written in function form as $D_{\tau}(\{F_i\}_{i=1}^{\tau},\{U_j\}_{j=1}^{\tau})$ that returns the set of helpers the newcomer has to access at time $\tau$, where $F_i$ and $U_j$ are the failed node at time $i$ and the set of unavailable nodes at time $j$, respectively. Since the helper selection function depends on the history of the failure/unavailability patterns and can change for each different time $\tau$, we term this scheme the \emph{dynamic helper selection (DHS)} scheme, for which the term ``dynamic'' emphasizes the time and history dependence of the helper selection rules. 

\par {\bf SHS:} A subset of the DHS schemes is the \emph{stationary helper selection (SHS)} schemes that assign fixed helper sets of $d$ nodes to each combination of a failed node and a set of unavailable nodes. Mathematically, the helper set decision at time $\tau$ in SHS can be written in function form as $D(F_{\tau},U_{\tau})$. The function $D(\cdot,\cdot)$ does not change with respect to the value of $\tau$ and the input arguments of $D(\cdot,\cdot)$ are $F_{\tau}$ and $U_{\tau}$, instead of the entire history $\{F_i\}_{i=1}^{\tau}$ and $\{U_j\}_{j=1}^{\tau}$. The idea is that, for a given node failure and a given set of unavailable nodes, the same helper set is used at any time instant. We can see that this construction is stationary because the helper sets do not change with time and only depend on current time $\tau$ failure and node unavailability information. 

\par All existing (non-blind) helper selection schemes can be interpreted as a form of SHS. For example, a popular way of helper selection when there are $r$ temporarily unavailable nodes \cite{prakash2012optimal,papailiopoulos2014locally,kamath2014codes,rawat2014optimal} is as follows. Each node $F$ is assigned a fixed set of $(d+r)$ candidate helper nodes. When node $F$ needs to be repaired, since at most $r$ nodes are temporarily unavailable, there are at least $d$ nodes that are still available in the candidate set. Then the newcomer arbitrarily contacts $d$ available nodes in the candidate helper set. Mathematically, such a scheme can be interpreted as an SHS scheme in the following way. Denote $\overline{D}(F_{\tau})$ as the set of $(d+r)$ candidate helpers of the failed node $F_{\tau}$ at time $\tau$. Again let $U_{\tau}$ denote the collection of temporarily unavailable nodes. Without loss of generality, we assume that $U_{\tau}\subseteq \overline{D}(F_{\tau})$ and $|U_{\tau}|=r$. Namely, there are exactly $r$ unavailable nodes and all of them are within the candidate set $\overline{D}(F_{\tau})$. This is possible since any scheme has to consider the worst case\footnote{A more rigorous formulation has to take an adversarial approach. Namely, if $U_{\tau}$ is not completely inside $\overline{D}(F_{\tau})$ or $|U_{\tau}|<r$, then we simply let an adversary to choose a new $U'_{\tau}$ satisfying $(U_{\tau}\cap \overline{D}(F_{\tau}))\subseteq U'_{\tau}\subseteq \overline{D}(F_{\tau})$ and $|U'_{\tau}|=r$.} scenario, in which all $r$ unavailable nodes are within $\overline{D}(F_{\tau})$. Then, we simply set the SHS function $D(F_{\tau},U_{\tau})$ by 

\begin{align}
D(F_{\tau},U_{\tau})=\overline{D}(F_{\tau})\backslash U_{\tau}.\label{eq:special-SHS}
\end{align}
Since $\overline{D}(F_{\tau})$ has $(d+r)$ nodes and $|U_{\tau}|=r$, the function $D(F_{\tau},U_{\tau})$ indeed returns $d$ helpers for the given $(F_\tau,U_\tau)$.

\par {\bf BHS:} The last type of helper selection, which is the most basic, is \emph{blind helper selection (BHS)} that allows the newcomer to access any arbitrarily selected $d$ nodes of the surviving nodes. This scheme was initially assumed for RCs in \cite{dimakis2010network}.

\subsection{Helper Selection Schemes In Existing Works}\label{subsec:helper_existing}

\par The helper selection schemes of all existing LRC constructions are SHS. Specifically, for $r=0$ (i.e., nodes are always available), LRCs \cite{gopalan2012locality,prakash2012optimal,papailiopoulos2014locally} use SHS where each node is assigned a fixed set of $d$ helper nodes. For $r>0$, LRCs \cite{kamath2014codes,rawat2014optimal} assign each node a fixed set of $(d+r)$ helper nodes and during repair the newcomer can arbitrarily connect to any $d$ nodes of the $(d+r)$ nodes in its helper set. As explained in Section~\ref{subsec:types_schemes}, such a scheme is a special form of SHS. Almost all LRCs considered in the existing literature use the above method to handle temporary node unavailability. To our knowledge, the only example in the literature that does not use the above helper selection method is in \cite{pamies2013locally}, for which the helper selection is based on a carefully designed $D(F_{\tau},U_{\tau})$ instead of \eqref{eq:special-SHS}.

\section{Information Flow Graphs and the Corresponding Graph-Based Analysis}\label{sec:ifg}
Before introducing our main results, we quickly explain the concepts of information flow graphs (IFGs) and the corresponding analysis, which was first introduced in \cite{dimakis2010network}. For readers who are not familiar with IFGs, we provide its detailed description in Appendix~\ref{app:ifg}.

\par Intuitively, each IFG reflects one unique history of the failure patterns and the helper selection choices from time $1$ to $(\tau-1)$ \cite{dimakis2010network}. Consider any given helper selection scheme $A$ which can be either DHS or SHS. Since there are infinitely many different failure patterns $F_{\tau}$ and infinitely many different unavailable node sets $U_{\tau}$ (since we consider $\tau=1$ to $\infty$), there are infinitely many IFGs corresponding to the same given helper selection scheme $A$ since the IFG grows according to the helper choices $D_{\tau}(\{F_i\}_{i=1}^{\tau},\{U_j\}_{j=1}^{\tau})$ for DHS or $D(F_{\tau},U_{\tau})$ for SHS. We denote the collection of all possible IFGs of a given helper selection scheme $A$ by $\mathcal{G}_A(n,k,d,r,\alpha,\beta)$. We define $\mathcal{G}(n,k,d,r,\alpha,\beta)=\bigcup_{\forall A}\mathcal{G}_A(n,k,d,r,\alpha,\beta)$ as the union over all possible helper selection schemes $A$. We sometimes drop the input argument and use $\mathcal{G}_A$ and $\mathcal{G}$ as shorthands. The collection $\mathcal{G}$ can also be viewed as the IFGs generated by BHS. The reason is that BHS blindly selects the helpers and thus will take into consideration {\em all possible ways} of growing the IFG. As a result, $\mathcal{G}_\text{BHS}=\mathcal{G}=\bigcup_{\forall A}\mathcal{G}_A(n,k,d,r,\alpha,\beta)$.

\par Given an IFG $G\in \mathcal{G}$, we use $\DC(G)$ to denote the collection of all ${n\choose k}$ {\em data collector nodes} in $G$ \cite{dimakis2010network}. Each data collector $t\in\DC(G)$ represents one unique way of choosing $k$ out of $n$ active nodes when reconstructing the file. Given an instance of the IFGs $G\in\mathcal{G}$ and a data collector $t\in \DC(G)$, we use $\mincut_G(s,t)$ to denote the {\em minimum cut value} \cite{west2001introduction} separating $s$, the root node (source node) of $G$, and $t$.

\par For any helper scheme $A$ and given system parameters $(n,k,d,r,\alpha,\beta)$, the results in \cite{ahlswede2000network} prove that the following condition is \emph{necessary} for the existence of any distributed storage network with helper selection scheme $A$ that can meet the design requirement $(n,k,d,r,\alpha,\beta)$:
\begin{align}
\min_{G\in \mathcal{G}_A}\min_{t\in \DC(G)}\mincut_G(s,t)\geq \mathcal{M}. \label{eq:condition}
\end{align}
If we limit our focus to a distributed storage network with BHS, then the above necessary condition becomes
\begin{align}
\min_{G\in \mathcal{G}}\min_{t\in \DC(G)}\mincut_G(s,t)\geq \mathcal{M}.\label{eq:condition-BR}
\end{align}
Reference \cite{dimakis2010network} later found a closed-form expression of the left-hand side (LHS) of \eqref{eq:condition-BR}
\begin{align}
\min_{G\in \mathcal{G}}\min_{t\in \DC(G)}\mincut_G(s,t)=\sum_{i=0}^{k-1} \min((d-i)^+\beta,\alpha)\label{eq:ex_low_b},
\end{align}
where $(x)^+=\max(x,0)$, which allows us to numerically check whether \eqref{eq:condition-BR}$\geq$$\mathcal{M}$ for any $(n,k,d,r,\alpha,\beta)$ values. Specifically, the necessary condition \eqref{eq:condition-BR} becomes 
\begin{align}
\sum_{i=0}^{k-1} \min((d-i)^+\beta,\alpha)\geq \mathcal{M}.\label{eq:Dimakis}
\end{align}

\par Reference \cite{wu2010existence} further proves that when considering a fixed but sufficiently large finite field $\GF(q)$, \eqref{eq:Dimakis} is not only necessary but also sufficient for the existence of a BHS-based distributed storage network that meets the design requirement $(n,k,d,r,\alpha,\beta)$.

Fix the values of $(n,k,d,r)$, two points on a storage-bandwidth tradeoff curve of any given helper selection scheme $A$ are of special interest: the minimum bandwidth regenerating (MBR) and minimum storage regenerating (MSR) points. These points can be defined as follows:
\begin{definition} For any given $(n,k,d,r)$ values, the MBR point $(\alpha_{\text{MBR}}, \beta_{\text{MBR}})$ of a helper scheme $A$ is defined by
\begin{align}
\beta_{\text{MBR}}&\stackrel{\Delta}{=}\min \{\beta: (\alpha,\beta)\text{ satisfies \eqref{eq:condition} and } \alpha=\infty\}\label{eq:MBR-beta-def}\\
\alpha_{\text{MBR}}&\stackrel{\Delta}{=}\min \{\alpha: (\alpha,\beta)\text{ satisfies \eqref{eq:condition} and } \beta=\beta_{\text{MBR}}\}.\label{eq:MBR-alpha-def}
\end{align}
\end{definition}

\begin{definition} For any given $(n,k,d,r)$ values, the MSR point $(\alpha_\text{MSR}, \beta_{\text{MSR}})$ of a helper scheme $A$ is defined by
\begin{align}
\alpha_{\text{MSR}}&\stackrel{\Delta}{=}\min \{\alpha: (\alpha,\beta)\text{ satisfies \eqref{eq:condition} and } \beta=\infty\}\label{eq:MSR-def}\\
\beta_{\text{MSR}}&\stackrel{\Delta}{=}\min \{\beta: (\alpha,\beta)\text{ satisfies \eqref{eq:condition} and } \alpha=\alpha_{\text{MSR}}\}.\nonumber
\end{align}
\end{definition}
Specifically, the MBR and MSR points are the two extreme ends\footnote{An alternative definition of the MSR point is when each node only stores $\alpha'_{\text{MSR}}=\frac{\mathcal{M}}{k}$ packets. The difference between these two definitions is as follows. The $\alpha_{\text{MSR}}$ in \eqref{eq:MSR-def} is the smallest possible storage under a given helper selection scheme $A$ and given reliability requirement $(n,k,d,r)$. In contrast, the alternative definition $\alpha'_{\text{MSR}}=\frac{\mathcal{M}}{k}$ is the smallest possible storage that can be achieved by an $(n,k)$ erasure code, which always requests repair data from $k$ helpers. For example, when $(n,k,d,r)=(5,3,2,1)$, one can prove that regardless how one design the helper selection scheme $A$, we always have $\alpha_{\text{MSR}}\geq \frac{\mathcal{M}}{2}$. Namely, the smallest achievable storage $\alpha_{\text{MSR}}$ is lower bounded by $\frac{\mathcal{M}}{2}$. Since no scheme can possibly achieve $\alpha'_{\text{MSR}}=\frac{\mathcal{M}}{3}$, the alternative MSR definition will say that the MSR point is not achievable for the parameter $(n,k,d,r)=(5,3,2,1)$} of the bandwidth-storage tradeoff curve in \eqref{eq:condition}.

\par The above graph-based analysis also allows us to define the optimality of different helper selection schemes. 
\begin{definition} \label{def:scheme-optimal}
For any given $(n,k,d,r)$ values, a helper selection scheme $A$ is \emph{optimal}, if for any DHS scheme $B$ the following is true
\begin{align}
\min_{G\in \mathcal{G}_A}\min_{t\in \DC(G)}\mincut_G(s,t)\geq \min_{G\in \mathcal{G}_B}\min_{t\in \DC(G)}\mincut_G(s,t)\nonumber
\end{align}
for all $(\alpha,\beta)$ combinations. That is, scheme $A$ has the best $(\alpha,\beta)$ tradeoff curve among all DHS schemes and thus allows for the protection of the largest possible file size.
\end{definition}

\section{The Main Results}\label{sec:main_results}
The main contributions of this work are the answers to the following two questions. Question 1: When designing an optimal helper selection scheme, is it sufficient\footnote{A simple analogy is as follows. It is known that for binary symmetric channels linear codes are capacity-achieving. Namely there is no need to search for non-linear codes. When considering network coding, again linear codes are capacity-achieving for the single multicast setting. But the seminal results in \cite{dougherty2005insufficiency} prove that linear codes are not sufficient for the multiple unicast setting. For this work, we would like to answer the question whether SHS is sufficient (capacity-achieving) for all $(n,k,d,r)$ values?} to limit the search scope to only considering SHS schemes? Question 2: We observe that for some $(n,k,d,r)$ values, even the best DHS/SHS schemes do not do better than the simplest BHS scheme. We call such $(n,k,d,r)$ values being {\em indfferent to helper selection} since for those $(n,k,d,r)$ the BHS is as good as any schemes. The question to be answered is thus for any arbitrarily given $(n,k,d,r)$, is there any way to quickly check whether it is {\em indifferent to helper selection} or not? 

\par We answer the first question in the following Propositions~\ref{prop:result_SHS} and \ref{prop:result_DHS} and answer partially the second question in Propositions~\ref{prop:result_conditions} to \ref{prop:new3}.

\begin{proposition}\label{prop:result_SHS}
For $(n,k,d,r)=(5,3,2,1)$ and $(5,4,2,1)$, and any arbitrary $(\alpha,\beta)$ values, there exists no SHS scheme that can protect a file of size larger than that of BHS.
\end{proposition}

\begin{proposition}\label{prop:result_DHS}
For $(n,k,d,r)=(5,3,2,1)$ and $(5,4,2,1)$, there exists a pair of $(\alpha,\beta)$ values such that one can find a DHS scheme that can protect a file of size strictly larger than that of BHS. Furthermore, we explicitly devise a DHS scheme that is provably optimal for $(n,k,d,r)=(5,3,2,1)$ and $(5,4,2,1)$. See Definition~\ref{def:scheme-optimal}. 
\end{proposition}

\par We can see, by Proposition~\ref{prop:result_DHS}, that using DHS we can protect a file size strictly larger than that of the best SHS scheme. This answers Question 1 by showing that SHS is not enough to achieve the optimal performance for $(n,k,d,r)=(5,3,2,1)$ and $(5,4,2,1)$. At least for these two parameter values, a DHS scheme is necessary. A byproduct of our optimal DHS scheme is that it achieves the MBR and MSR points simultaneously. Specifically, it simultaneously minimizes the bandwidth and storage for $(n,k,d,r)=(5,3,2,1)$ and $(5,4,2,1)$.

\par The following proposition answers the second question by providing conditions that can be used to check whether a given $(n,k,d,r)$ value is indifferent to helper selection or not.

\begin{proposition}\label{prop:result_conditions}
If the following inequality
\begin{align}
k\leq \left\lceil\frac{n-r}{n-d-r}\right\rceil\label{eq:ccw-cond1}
\end{align}
holds, then for any arbitrary $(\alpha,\beta)$ values there exists no DHS scheme that can protect a file of size larger than that of BHS.

\par If the following inequality
\begin{align}
\min(d+1,k)>\left\lceil \frac{n}{n-d-r} \right\rceil\label{eq:ccw-cond2}
\end{align} 
holds, then there exists an SHS scheme and a pair of $(\alpha,\beta)$ such that we can protect a file of size strictly larger than that of BHS.
\end{proposition}

\par There are some $(n,k,d,r)$ values that satisfy neither \eqref{eq:ccw-cond1} nor \eqref{eq:ccw-cond2} for which it remains open whether those $(n,k,d,r)$ are indifferent to helper selection or not. Therefore the characterization in Proposition~\ref{prop:result_conditions} is not tight. 

\par For the cases\footnote{Arguably, the cases of small $r$ are more interesting from a practical perspective.} of $r\leq 1$, we can further sharpen the results as follows. 
\begin{proposition}\cite[Proposition~1]{arxiv1}\label{prop:new1}
For any $(n,k,d,r)$ values satisfying $r=0$, if either \eqref{eq:ccw-cond1} or 
\begin{align}
d=1,~k=3,~\text{and $n$ is odd}\label{eq:ccw-cond3}
\end{align}
holds, then for any $(\alpha,\beta)$ values, there exists no DHS scheme that can protect a file of size larger than that of BHS. If neither \eqref{eq:ccw-cond1} nor \eqref{eq:ccw-cond3} holds, then there exists an SHS scheme and a pair of $(\alpha,\beta)$ such that we can protect a file of size strictly larger than that of BHS.
\end{proposition}
\begin{proposition}\label{prop:new2}
For any $(n,k,d,r)$ values satisfying $r=1,d=1$, if either (i) \eqref{eq:ccw-cond1} holds or (ii) $k=3$ or (iii)
\begin{align}
k=4,~\text{and}~ n\bmod 3\neq 0
\end{align}
holds, then for any $(\alpha,\beta)$ values, there exists no DHS scheme that can protect a file of size larger than that of BHS. If none of (i)-(iii) holds, then there exists an SHS scheme and a pair of $(\alpha,\beta)$ such that we can protect a file of size strictly larger than that of BHS.
\end{proposition}

\begin{proposition}\label{prop:new3}
For any $(n,k,d,r)$ value satisfying $r=1,d=2$, if \eqref{eq:ccw-cond1} does not hold then there exists a DHS scheme and a pair of $(\alpha,\beta)$ such that we can protect a file of size strictly larger than that of BHS.
\end{proposition}
\par Propositions~\ref{prop:new1} to \ref{prop:new3} close the gap in Proposition~\ref{prop:result_conditions} and provide tight characterization for the cases of ``$r=0$'' and  ``$r=1, d\leq 2$.'' Propositions~\ref{prop:result_conditions} to \ref{prop:new3} quickly leads to the following corollary.
\begin{corollary} For any $(n,k,d,r)$ satisfying $r\leq 1$, $d\leq 5$, and 
\begin{align}
(n,k,d,r)\notin \{(7,3,3,1), (9,3,4,1), (7,4,4,1), (11,3,5,1)\},
\end{align}
we can easily determine whether $(n,k,d,r)$ is indifferent to helper selection or not by checking some very simple conditions. 
\end{corollary}

\par The proofs of Propositions~\ref{prop:result_SHS} and \ref{prop:result_DHS} will be presented in Section~\ref{sec:dy_better}. The proofs of the converse and the achievability parts of Proposition~\ref{prop:result_conditions} will be provided in Sections~\ref{sec:mincut}. 

\par Proposition~\ref{prop:new1} focuses on the special case of $r=0$ and is a restatement of the results in \cite[Proposition~1]{arxiv1}. The proof of Proposition~\ref{prop:new2} is relegated to Appendix~\ref{app:new2}. We close this section by providing the proof of Proposition~\ref{prop:new3}, which reveals a connection between Proposition~\ref{prop:new3} and Propositions~\ref{prop:result_SHS} and \ref{prop:result_DHS}.

\begin{IEEEproof}[Proof of Proposition~\ref{prop:new3}]
By simple counting arguments provided in Appendix~\ref{app:gap}, one can show that among all $(n,k,d,r)$ satisfying $r=1$ and $d=2$, there are only two instances $(n,k,d,r)=(5,3,2,1)$ and $(5,4,2,1)$ that satisfy neither \eqref{eq:ccw-cond1} nor \eqref{eq:ccw-cond2}. Namely, any other $(n,k,d,r)$ satisfies at least one of \eqref{eq:ccw-cond1} and \eqref{eq:ccw-cond2}. By Proposition~\ref{prop:result_conditions} we only need to decide whether a given $(n,k,d,r)$ is indifferent to helper selection for these two instances. 

\par At the same time, by Propositions~\ref{prop:result_SHS} and \ref{prop:result_DHS} there exists a DHS scheme and a pair of $(\alpha,\beta)$ such that we can protect a file of size strictly larger than that of BHS. The proof is thus complete. 
\end{IEEEproof}

\section{Stationary Helper Selection Is Insufficient} \label{sec:dy_better}
\par The proofs of Propositions~\ref{prop:result_SHS} and \ref{prop:result_DHS} are provided in Sections~\ref{subsec:SHS-BHS_proof} and \ref{subsec:DHS-better}, respectively. Jointly, they prove that SHS is insufficient in terms of achieving the optimal repair BW. A byproduct of the results in Propositions~\ref{prop:result_SHS} and \ref{prop:result_DHS} is a simple proof showing that functional repair can be strictly better than exact repair, which is provided in Section~\ref{subsec:byproduct}. 

\subsection{Proof of Proposition~\ref{prop:result_SHS}} \label{subsec:SHS-BHS_proof}

\begin{figure}[h!]
\centering
\includegraphics[width=0.4\textwidth]{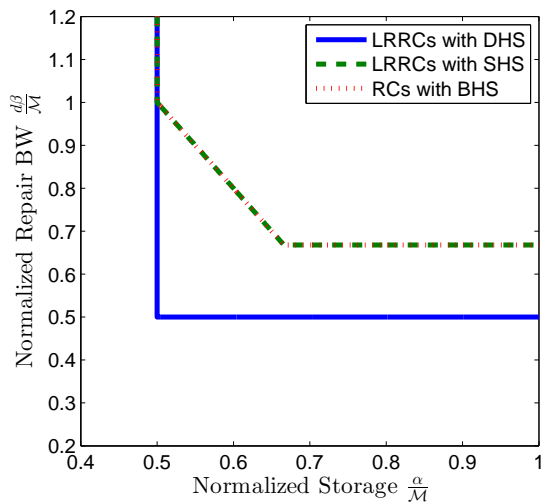}
\caption{Storage-bandwidth tradeoff curves of LRRCs with DHS, LRRCs with SHS, and RCs with BHS for $(n,k,d,r)=(5,3,2,1)$.}
\label{fig:storage_vs_bandwidth_(5-3-2)}
\end{figure}

\par We first consider the case of $(n,k,d,r)=(5,3,2,1)$. Since BHS is used, the newcomer can access any $d=2$ out of $3=n-r-1$ available nodes and it thus naturally handles node unavailability ($r=1$). The storage-BW tradeoff when BHS is used can then be derived directly from plugging in $(n,k,d)=(5,3,2)$ in \eqref{eq:Dimakis}. Namely, as long as the BHS policy is used, the storage-BW tradeoff curve must satisfy
\begin{align}
\min(2\beta,\alpha)+\min(\beta,\alpha)\geq \mathcal{M}\label{eq:BHS-tradeoff},
\end{align}
where $\mathcal{M}$ is the file size. A normalized storage-BW tradeoff curve of \eqref{eq:BHS-tradeoff}, is plotted in Fig.~\ref{fig:storage_vs_bandwidth_(5-3-2)}. Namely, if each node stores only half of the overall file $\frac{\alpha}{\mathcal{M}}=0.5$, then the normalized repair BW is $\frac{d\beta}{\mathcal{M}}=1$. However, if we are willing to use a larger normalized storage size $\frac{\alpha}{\mathcal{M}}=\frac{2}{3}$ rather than $\frac{1}{2}$, we can reduce the normalized BW $\frac{d\beta}{\mathcal{M}}$ from 1 to $\frac{2}{3}$. Note that when BHS is used, we are essentially analyzing the original RCs in \cite{dimakis2010network} with $(n,k,d)=(5,3,2)$. Therefore, the actual protection level satisfies $k^*=d=2$ which is strictly smaller than the target protection level $k=3$. This means that any code construction with BHS is overprotecting the data.

\par In the following, we will show that even if one is allowed to choose the helpers in an intelligent way (other than BHS), we are still not able to improve the storage-BW tradeoff curve in \eqref{eq:BHS-tradeoff} if we are restricted to using only SHS One implication of this result is that when $(n,k,d,r)=(5,3,2,1)$ any existing/future LRC scheme will have the same performance as the original RC if restricted to using only SHS schemes. 

\par Consider any SHS scheme $A$ with the corresponding stationary helper selection function being $D(F,U)$ where $F$ is the failed node and $U=\{j\}$ is the set containing the temporarily unavailable node $j$ since we now focus on $r=1$. For simplicity, we sometimes just say node $U$ is unavailable when it is clear from the context that $|U|=r=1$. Recall that $\mathcal{G}_A$ is the collection of IFGs that are grown according to the helper selection scheme $A$. The main idea is to prove that no matter how we design the $D(F,U)$, we are bound to have a graph $G^*\in \mathcal{G}_A$ for which $\min_{t\in \DC(G^*)}\mincut_{G^*}(s,t)$ equals to the LHS of \eqref{eq:BHS-tradeoff}. This implies that
\begin{align}
\min_{G\in \mathcal{G}_A} \min_{t\in \DC(G)} \mincut_{G}(s,t)\leq \min(2\beta,\alpha)+\min(\beta,\alpha).\label{new2}
\end{align}
By the fact that $G_A\subseteq\mathcal{G}$, and by \eqref{eq:ex_low_b}, we thus have
\begin{align}
\min_{G\in \mathcal{G}} \min_{t\in \DC(G)} \mincut_{G}(s,t)&\leq \min_{G\in \mathcal{G}_A} \min_{t\in \DC(G)} \mincut_{G}(s,t)\nonumber\\
&\leq \min(2\beta,\alpha)+\min(\beta,\alpha)\nonumber\\
&= \min_{G\in \mathcal{G}} \min_{t\in \DC(G)} \mincut_{G}(s,t).\label{eq:ccw-new4}
\end{align}

As a result all the inequalities in \eqref{eq:ccw-new4} must be equality. The storage-BW tradeoff curve of helper scheme $A$ in \eqref{eq:condition} is thus identical to the RC tradeoff in \eqref{eq:BHS-tradeoff}.

\par To that end, we will prove that, regardless how we design the helper selection $D(F,U)$ function, we can always find a graph $G^*\in \mathcal{G}_A$ such that there exist 3 active nodes $x$, $y$, and $z$ satisfying (i) each node has been repaired at least once, and (ii) $x$ is a helper when repairing $y$, and (iii) both $x$ and $y$ are the helpers when repairing $z$. Considering the cut in $G^*$ that directly separates the source (root) from $\{x,y,z\}$, we can observe that node $x$ will contribute $\min(2\beta,\alpha)$ to the cut value; node $y$ will contribute $\min(\beta,\alpha)$ to the cut value since node $x$ was the helper of node $y$; and node $z$ will contribute 0 to the cut value since both $x$ and $y$ are the helpers of $z$. Therefore, the cut that separates the source (root) directly from $\{x,y,z\}$ will have the cut-value being $\min(2\beta,\alpha)+ \min(\beta,\alpha)$. As a result, the min-cut value $\mincut_{G^*}(s,\{x,y,z\})$ is no larger than the LHS of \eqref{eq:BHS-tradeoff}. We have thus proved \eqref{new2}.

\par We prove the existence of such an IFG $G^*\in\mathcal{G}_A$ by contradiction. Without loss of generality, suppose that $D(1,\{4\})=\{2,3\}$ in the SHS scheme $A$. Namely, if node 1 fails and node 4 is not available, then the newcomer (node 1) will access nodes 2 and 3 as helpers. This assumption can always be made true by relabeling the nodes. We consider the following 2 cases.

\par \emph{Case 1:} $D(2,\{1\})\neq \{4,5\}$, or $D(2,\{4\})\neq \{1,5\}$. Consider the following three subcases. Case 1.1: $D(2,\{1\})\neq \{4,5\}$. Since $D(2,\{1\})$, by definition, returns a subset of $\{1,2,3,4,5\}\backslash (\{F\}\cup U)=\{3,4,5\}$, we must have either $D(2,\{1\})=\{3,4\}$ or $\{3,5\}$. We now fail node 3 first and repair it following scheme $A$. (What is the rule that scheme $A$ uses to repair node 3 is irrelevant in our proof.) Then, fail node 2 and suppose node 1 is unavailable. Since $D(2,\{1\})=\{3,4\}$ or $\{3,5\}$ in Case 1.1, node 3 will definitely be a helper of node 2. Next, fail node 1 and assume node 4 is unavailable. Since $D(1,\{4\})=\{2,3\}$, node 1 will access nodes 2 and 3 for repair. We can observe that we have constructed such a $G^*$ where nodes $(x,y,z)=(3,2,1)$ satisfy properties (i) to (iii) in the previous paragraph. The proof for Case 1.1 is complete.

\par Case 1.2: $D(2,\{4\})\neq \{1,5\}$. Since $D(2,\{4\})$ returns a subset of $\{1,3,5\}$, we must either have $D(2,\{4\})=\{1,3\}$ or $\{3,5\}$. Similar to Case 1.1, we fail node 3 first. Then, we fail node 2 and assume that node 4 is unavailable. Since $D(2,\{4\})=\{1,3\}$ or $\{3,5\}$, node 3 must be a helper of node 2. Finally, fail node 1 and assume that node 4 is unavailable. Since $D(1,\{4\})=\{2,3\}$, node 1 will access nodes 2 and 3 for repair. In the end, nodes $(x,y,z)=(3,2,1)$ satisfy (i) to (iii).

\par \emph{Case 2:} $D(2,\{1\})=\{4,5\}$ and $D(2,\{4\})= \{1,5\}$. We consider two subcases. Case 2.1: $D(1,\{3\})\neq \{4,5\}$. Therefore, we must have $D(1,\{3\})=\{2,4\}$ or $\{2,5\}$. For ease of exposition, we say $D(1,\{3\})=\{2, v\}$ where $v$ is either node 4 or node 5. We now fail node $v$ first and repair it under scheme $A$. (What is the rule that scheme $A$ uses to repair node $v$ is irrelevant in our proof.) We then fail node 2 and assume that node 1 is unavailable. Since $D(2,\{1\})=\{4,5\}$, nodes 4 and 5 are the helpers of node 2. Then, fail node 1 and assume node 3 is unavailable. Since $D(1,\{3\})=\{2,v\}$, nodes 2 and $v$ are the helpers of node 1. Observe that $(x,y,z)=(v,2,1)$ satisfy (i) to (iii) and we have thus constructed such a $G^*$.

\par Case 2.2: $D(1,\{3\})=\{4,5\}$. We fail node 5 first and repair it under scheme $A$. (What is the rule that scheme $A$ uses to repair node $v$ is irrelevant in our proof.) We then fail node 1 while assuming that node 3 is unavailable. Since $D(1,\{3\})=\{4,5\}$, nodes 4 and 5 are the helpers of node 1. Then, fail node 2 and assume that node 4 is unavailable. Since in Case~2 we have $D(2,\{4\})= \{1,5\}$, nodes 5 and 1 are the helpers of node 2. Observe that $(x,y,z)=(5,1,2)$ satisfy (i) to (iii) and we have thus constructed such a $G^*$. 

\par Thus far, we have completed the proof for the case of $(n,k,d,r)=(5,3,2,1)$. We now discuss how to prove the case of $(n,k,d,r)=(5,4,2,1)$. By \eqref{eq:Dimakis}, one can directly prove that when BHS is used, the storage-BW tradeoff curve of $(n,k,d,r)=(5,4,2,1)$ is also governed by \eqref{eq:BHS-tradeoff}. 

\par To prove that the storage-BW tradeoff curve of SHS is also \eqref{eq:BHS-tradeoff}, we will prove that regardless how we choose the helper selection function $D(F,U)$, there always exists a graph $G^{**}\in \mathcal{G}_A$ such that there exists 4 active nodes $x$, $y$, $z$, and $w$ satisfying (i) each node has been repaired at least once, and (ii) $x$ is a helper when repairing $y$, (iii) both $x$ and $y$ are the helpers when repairing $z$, and (iv) the helper nodes of $w$ are a subset of $\{x,y,z\}$.

\par By the discussion in the previous proof of $(n,k,d,r)=(5,3,2,1)$, we can always find a $G^*$ such that there exist three active nodes $(x,y,z)$ satisfying (i) to (iii). Without loss of generality, assume the three active nodes are $(x,y,z)=(1,2,3)$. Then we fail node 4 and assume node 5 is unavailable. Since there are only 3 remaining nodes $\{1,2,3\}$, regardless how we choose $D(4,\{5\})$, the helpers of node 4 must be a subset of $\{1,2,3\}$. We call the IFG after repairing node 4, $G^{**}$. Choose $w=4$. Then nodes $(x,y,z,w)=(1,2,3,4)$ satisfy (i) to (iv). 

\par By similar arguments, one can easily check that the min-cut separating the root of $G^{**}$ and the four nodes $(x,y,z,w)$ is at most $\min(2\beta,\alpha)+ \min(\beta,\alpha)$. Therefore, the storage-BW tradeoff curve for any SHS scheme with $(n,k,d,r)=(5,4,2,1)$ must again be \eqref{eq:BHS-tradeoff}. The proof of Proposition~\ref{prop:result_SHS} is thus complete.

\subsection{Proof of Proposition~\ref{prop:result_DHS}}\label{subsec:DHS-better}
\par We first consider $(n,k,d,r)=(5,3,2,1)$ and prove that there exists a DHS scheme that has the following new strictly better tradeoff curve:
\begin{align}
2\min(2\beta,\alpha) \geq \mathcal{M}\label{eq:new-tradeoff},
\end{align}
i.e., it strictly outperforms the best possible SHS scheme, for which the tradeoff is governed by \eqref{eq:BHS-tradeoff}. Our proof is by explicit code construction with $\beta=1$, $\alpha=2$, and $\mathcal{M}=4$, which achieves the corner point of \eqref{eq:new-tradeoff}, also see Fig.~\ref{fig:storage_vs_bandwidth_(5-3-2)}. The scheme consists of two parts. {\bf Part I:} How to choose the helper nodes for a newcomer? {\bf Part II:} What is the coded data sent by each helper\footnote{Since $\alpha=2=d\beta$, each node simply stores all the $d\beta$ packets it has received in its local memory.} after the helpers are decided?

\par To describe {\bf Part I}, we need the following notation. We say node $i$ is the {\em parent} of node $j$ if (i) node $i$ was the helper of node $j$, and (ii) node $i$ has not been repaired since the failure of node $j$. For example, say node 1 fails and accesses nodes 2 and 3 as helpers. Then node 2 fails and accesses nodes 3 and 4. After the above two repairs, node 3 is a parent of node 1 but node 2 is not since node 2 has been repaired. On the other hand, both nodes 3 and 4 are parents of node 2.

\par The main idea of the proposed DHS scheme is to choose helpers such that no 3 nodes ever form a ``triangle''. Namely, we carefully choose the helpers of the newcomers so that we can avoid the existence of 3 nodes $\{x,y,z\}$ such that $x$ is the parent of both $y$ and $z$; and $y$ is the parent of $z$. We term this DHS scheme \emph{the Clique-Avoiding (CA) scheme}.

\par We now prove by induction that CA is always possible. In the beginning, all nodes are intact and no node is the parent of another. Therefore, there does not exist any 3 nodes forming a triangle. Suppose there is no triangle after $(\tau_0-1)$ repairs. At time $\tau=\tau_0$, suppose a node fails. 
For future reference, denote that node as node $z$. Since the network had no triangle at time $(\tau_0-1)$, we only need to ensure that the newcomer $z$ does not participate in any triangle {\em after} the repair. Denote the helper choice of the CA scheme for time $\tau=\tau_0$ by $\{x,y\}$. Therefore, we only need to carefully choose the helper set $\{x,y\}$ such that neither nodes $\{x,y,z\}$ nor nodes $\{y,x,z\}$ form a triangle after repair.

\par To prove the existence of such $\{x,y\}$, we observe that out of the $n-1=4$ surviving nodes, at most $r=1$ node is unavailable. As a result, the newcomer $c$ has 3 nodes to choose $d=2$ helpers from. Say, the nodes to choose from are $\{i,j,k\}$ and without loss of generality assume node $i$ is the oldest (being repaired the earliest) and node $k$ is the youngest (being repaired the latest) of the three. We argue that among the three pairs $\{(i,j), (j,k), (i,k)\}$ one of them must not form a parent-child pair. If not, i.e., all three are parent-child pairs, then nodes $\{i,j,k\}$ form a triangle in time $(\tau_0-1)$, which leads to contradiction. Say node $i$ is not a parent of $k$. Then, we choose nodes $i$ and $k$ to be the helper set $\{x,y\}$. As a result, neither nodes $\{x,y,z\}$ nor nodes $\{y,x,z\}$ form a triangle\footnote{Recall that node $i$ is older than node $k$ so node $k$ can never be a parent of node $i$.}. By induction CA is always possible.

\par Note that the CA scheme needs to use the {\em repair history} to decide which of the three pairs $\{(i,j), (j,k), (i,k)\}$ is not a parent-child pair and then chooses that pair as the helpers. Therefore, the choice of the helper sets may vary from time to time. This is a significant departure from the principle of associating each node $x$ with a {\em fixed helper set}. Because the CA scheme has to {\em dynamically} select the helpers based on repair history, we can see that the CA scheme is indeed a DHS scheme.

\par
\begin{figure}[h!]
\centering
\includegraphics[width=0.5\textwidth]{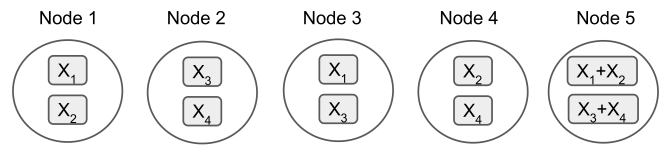}
\caption{The Code of the CA Scheme After Initialization.}
\label{fig:initialization}
\end{figure}

 We now describe {\bf Part II}: What is the coded data sent by each helper? Our construction uses only the {\em binary field} rather than high-order $\GF(q)$. A concrete example will be given after we give a complete description of our coding scheme. {\bf Initialization:} Recall that $\alpha=2$, $\beta=1$, and $\mathcal{M}=4$. Consider a file of 4 packets $X_1$ to $X_4$. Initially, we let nodes 1 and 2 store $\{X_1,X_2\}$ and $\{X_3, X_4\}$, respectively. We then let nodes 3 and 4 store packets $\{X_1, X_3\}$ and $\{X_2,X_4\}$, respectively. Finally, let node 5 store coded packets $\{[X_1+X_2], [X_3+X_4]\}$. The initialization phase is now complete. See Fig.~\ref{fig:initialization} for the illustration after the initialization phase.  

\par For easier description of our code construction, right after initialization, we artificially define nodes 1 and 2 as the parents of node 3 even though nodes 1 and 2 are not helpers of node 3. The reason is that the packets in node $3$ are $\{X_1,X_3\}$ and they can be viewed as if node 3 has failed and got repaired from nodes 1 and 2. See Fig.~\ref{fig:initialization} for illustration. Similarly, we artificially define nodes 1 and 2 as the parents of node 4 (resp.\ node 5) even though nodes 1 and 2 are not helpers of node 4 (resp.\ node 5).\footnote{Even with the artificially defined parent-child relationship, there is no triangle after initialization. We can thus use the same induction proof to show that CA is always possible after initialization.}

\par {\bf The regular repair operations:} Suppose node $a$ fails and one
other node is temporarily unavailable at time $\tau$. We run the aforementioned CA helper selection
scheme to find the helpers $b$ and $c$ for node $a$. Denote the two non-helper nodes by
$d$ and $e$. Each of $b$ and $c$ will send 1 packet to $a$ since $\beta=1$.
The packets are constructed as follows. {\bf Step 1:} Denote the two
(potentially coded) packets stored in $b$ by $Y_1^{(b)}$ and $Y_2^{(b)}$.
Among the three candidate packets $Y_1^{(b)}$, $Y_2^{(b)}$, and the binary sum
$[Y_1^{(b)}+Y_2^{(b)}]$, node $b$ will choose one packet, call it $Z_b^*$, and send it to $a$. 

\par Before describing how to choose $Z_b^*$, we construct two conditions based on the packets currently stored in nodes $c$, $d$, and $e$. If nodes $c$ and $d$ jointly contain 4 {\em linearly
independent} packets, then we construct Condition~1 to be ``$ Z_b^*$ cannot be expressed as a linear combination of the two packets stored in $d$.'' Otherwise, we construct Condition~1 to be ``$ Z_b^*$ cannot be expressed as a linear combination of the packets
stored in $c$ and $d$.'' Namely, depending on the coded packets stored in nodes $c$ and $d$, Condition~1 can be one of the above two different statements. Similarly, if nodes $c$ and $e$ jointly contain 4 linearly independent packets, then we construct Condition~2 to be ``$Z_b^*$ cannot be expressed as a
linear combination of the two packets stored in $e$.'' Otherwise, we construct Condition~2 to be ``$Z_b^*$ cannot be expressed as a linear combination of the packets
stored in $c$ and $e$.''

\par After constructing the two conditions, we require the choice $Z_b^*$ to satisfy simultaneously both Conditions 1 and 2. If there is more than one choice of $Z_b^*$ satisfying both conditions, then an arbitrary one of those $Z_b^*$ will do.

\par {\bf Step 2:} Denote the two packets stored in $c$ by $Y_1^{(c)}$ and $Y_2^{(c)}$. Among three candidate packets $Y_1^{(c)}$, $Y_2^{(c)}$, and the binary sum $[Y_1^{(c)}+Y_2^{(c)}]$, node $c$ will choose one packet, call it $Z_c^*$, and send it to node $a$. We require the packet $Z_c^*$ to satisfy simultaneously: (i) $ Z_c^*$ cannot be expressed as a linear combination of $Z_b^*$ and the two packets stored in $d$; and (ii) $ Z_c^*$ cannot be expressed as a linear combination of $Z_b^*$ and the two packets stored in $e$.

\par Once node $a$ receives $Z_b^*$ and $Z_c^*$, it stores both packets in its local memory (since $\alpha=2$). 

\begin{lemma}[feasibility of the proposed scheme] \label{lem:feasibility_ca} We can always find the $Z_b^*$ and $Z_c^*$ satisfying the specified conditions. As a result the code can be iteratively constructed for all time $\tau=1$ to $\infty$. 
\end{lemma}
\par The proof of Lemma~\ref{lem:feasibility_ca} is relegated to Appendix~\ref{app:ca_existence}. 

\begin{proposition}\label{prop:code-existence}
Using the above construction (Parts~I and II), for any time $\tau$, any $k=3$ nodes can always reconstruct the original $n=4$ packets
$X_1$ to $X_4$. Such a binary code construction
$(\alpha,\beta,\mathcal{M})=(2,1,4)$ thus satisfies the reliability requirement $(n,k,d,r)=(5,3,2,1)$.
\end{proposition}

The proof of Proposition~\ref{prop:code-existence} is relegated to Appendix~\ref{app:ca_existence}.

\par Let us use an example to illustrate our construction. Suppose after initialization, node 3 fails and node 2 is unavailable. Newcomer 3 thus has to access two of the nodes $\{1,4,5\}$ for repair. Since node 1 is the parent of both nodes $4$ and $5$, the CA scheme will avoid choosing $\{1,4\}$ and $\{1,5\}$ and select helpers $\{4,5\}$ instead. Specifically, $a=3$, $b=4$, and $c=5$; and $d=1$ and $e=2$.

\par Since node $b=4$ stores $\{X_2,X_4\}$, see Fig.~\ref{fig:initialization}, the three candidates for $Z_b^*$ are $X_2$, $X_4$, and $[X_2+X_4]$. 

\par Since node $c=5$ stores  $\{[X_1+X_2],[X_3+X_4]\}$ and node $d=1$ stores $\{X_1,X_2\}$, these two thus contain 4 linearly independent packets. Condition~1 becomes ``$Z_b^*$ cannot be any linear expression of packets $X_1$ and $X_2$, the packets in node $d$.'' Similarly, since $e=2$ stores $\{X_3,X_4\}$ and jointly nodes $c$ and $e$ contain 4 linearly independent packets, Condition~2 becomes ``$Z_b^*$ cannot be any linear expression of packets $X_3$ and $X_4$, the packets in node $e$.'' Out of the three candidates  $X_2$, $X_4$, and $[X_2+X_4]$, only the coded packet $[X_2+X_4]$ can satisfy both conditions simultaneously. Therefore we choose $Z_b^*=[X_2+X_4]$. 

\par Since node $c=5$ stores  $\{[X_1+X_2],[X_3+X_4]\}$, the three candidates for $Z_c^*$ are $[X_1+X_2]$, $[X_3+X_4]$, and $[X_1+X_2+X_3+X_4]$. The choice of $Z_c^*$ thus has to satisfy simultaneously (i) $Z_c^*$ is not a linear combination of $Z^*_b=[X_2+X_4]$ and the two packets $X_1$ and $X_2$ in node $d$; and (ii)  $Z_c^*$ is not a linear combination of $Z^*_b=[X_2+X_4]$ and the two packets $X_3$ and $X_4$ in node $e$. Out of the three candidates  $[X_1+X_2]$, $[X_3+X_4]$, and $[X_1+X_2+X_3+X_4]$, only the coded packet $[X_1+X_2+X_3+X_4]$ can satisfy both conditions simultaneously. Therefore, we choose $Z_c^*= [X_1+X_2+X_3+X_4]$. 

\par In the end,  $Z_b^*=[X_2+X_4]$ will then be sent to node $a$ from node $b$ and $Z_c^*=[X_1+X_2+X_3+X_4]$ will be sent to node $a$ from node $c$. Newcomer $a$ will then store both packets in its storage. The same repair process can then be repeated and applied to any arbitrary next newcomer.

\par The above CA helper selection scheme (Part I) and its code construction (Part II) thus achieve the new storage-BW tradeoff in \eqref{eq:new-tradeoff} for the case of $(n,k,d,r)=(5,3,2,1)$. 

\par For the case of $(n,k,d,r)=(5,4,2,1)$ we notice that we can use the same code\footnote{In this way, we are {\em overprotecting} the data since the actual $k^*=3$ but the target $k=4$.} that is constructed for $(n,k,d,r)=(5,3,2,1)$ to achieve the same storage-BW tradeoff in \eqref{eq:new-tradeoff}.  In the end of Section~\ref{subsec:SHS-BHS_proof}, we have already proven that the best SHS storage-BW tradeoff curve of $(n,k,d,r)=(5,4,2,1)$ is governed by \eqref{eq:BHS-tradeoff}. As a result Proposition~\ref{prop:result_DHS} is proven for $(n,k,d,r)=(5,3,2,1)$ and $(5,4,2,1)$. 

\subsection{The Optimal Solution For $(n,k,d,r)=(5,3,2,1)$ and $(5,4,2,1)$}
In this subsection, we prove that there exists no scheme (DHS or SHS) that can outperform the storage-BW tradeoff curve in \eqref{eq:new-tradeoff}. Therefore, the scheme described in Section~\ref{subsec:DHS-better} is optimal. 

\begin{proposition} \label{prop:absolute-optimal}
Suppose $(n,k,d,r)=(5,3,2,1)$ or $(5,4,2,1)$ and consider any arbitrary DHS scheme $A$. We have that
\begin{align} \label{eq:dynamic_optimal}
\min_{G\in \mathcal{G}_A} \min_{t\in \DC(G)} \mincut_{G}(s,t)\leq 2\min(2\beta,\alpha).
\end{align}

\end{proposition}

\par Observe that, by \eqref{eq:new-tradeoff}, the proposed CA scheme and the corresponding code construction achieve the upper bound in Proposition~\ref{prop:absolute-optimal} above. Therefore, we have that the proposed scheme is indeed optimal and there exists no helper selection scheme that can outperform it.

\begin{IEEEproof}[Proof of Proposition~\ref{prop:absolute-optimal}]
We first prove this proposition for $(n,k,d,r)=(5,3,2,1)$. Consider an IFG $G^*\in \mathcal{G}_A$ such that all its nodes have been repaired before. Consider the newest node in $G^*$ that we denote by $z$. Observe that $z$ must be connected to two older active nodes, call them $x$ and $y$. Now, consider a data collector that is connected to $\{x,y,z\}$. We can see that node $x$ will contribute $\min(2\beta,\alpha)$ to the value of the cut that directly separates  the root  from the three nodes $\{x,y,z\}$. Moreover, node $y$ will contribute at most $\min(2\beta,\alpha)$ to the value of the cut that directly separates $\{x,y,z\}$. On the other hand, node $z$ cannot contribute anything to the cut-value since it is connected to both $x$ and $y$. Therefore, the value of the cut that directly separates the root and $\{x,y,z\}$ is at most $2\min(2\beta,\alpha)$. As a result, the minimum cut-value $\mincut_{G^*}(s,t)$ for that particular $t$ is upper bounded by $2\min(2\beta,\alpha)$. Taking the minimum of all possible $t$ and all possible $G\in \mathcal{G}_A$, we get the inequality in \eqref{eq:dynamic_optimal} and the proof is complete.

\par For the case of $(n,k,d,r)=(5,4,2,1)$, consider an IFG $G^*\in \mathcal{G}_A$ such that all its nodes have been repaired before. Consider the newest node in $G^*$ that we denote by $z$. Observe that $z$ must be connected to two older active nodes, call them $x$ and $y$. Denote the nodes other than nodes $x$, $y$, and $z$, by nodes $w$ and $u$. We fail node $w$ and make $u$ temporarily unavailable. Repair node $w$ according to the given scheme $A$, which must access 2 out of the three remaining nodes $\{x,y,z\}$. 

\par Now, consider a data collector that is connected to $\{x,y,z,w\}$. We can see that node $x$ will contribute $\min(2\beta,\alpha)$ to the value of the cut that directly separates the root from $\{x,y,z,w\}$ and node $y$ will contribute at most $\min(2\beta,\alpha)$ to the value of that cut. Nodes $z$ and $w$ will not contribute any amount to the cut value since $z$ is connected to $\{x,y\}$ and $w$ is connected to two of $\{x,y,z\}$. By the verbatim argument as in the case of $(n,k,d,r)=(5,3,2,1)$, we have thus proven \eqref{eq:dynamic_optimal} for the case of $(n,k,d,r)=(5,4,2,1)$. 
\end{IEEEproof}

\subsection{A Byproduct of Propositions~\ref{prop:result_SHS} and~\ref{prop:result_DHS}}\label{subsec:byproduct}

\par As we saw in Table~\ref{table1}, we have two repair modes in distributed storage codes: \emph{functional repair} and \emph{exact repair}. Recall that in functional repair, nodes are allowed to store any functions of the original data, i.e., nodes do not have to retain the same packets at all times. In exact repair, however, nodes are required to store the same packets at all times. RCs \cite{dimakis2010network} were originally proposed with functional repair since functional repair is more general and could potentially lead to more repair-BW reduction. Exact repair was subsequently considered as it was observed that it can decrease overhead compared to functional repair due to the fact that the decoding and repairing rules are fixed in exact repair as opposed to the changing rules in functional repair. Moreover, it is possible using an exact repair code to have the original data be the systematic packets of the code which greatly facilitates data retrieval and reconstruction. Exact repair codes that achieve the MSR point of RCs (with BHS) were given in \cite{shah2012interference,rashmi2011optimal,cadambe2013asymptotic} and for the MBR point in \cite{rashmi2011optimal,shah2012distributed}. In \cite{shah2012distributed}, it was shown that the majority of the interior points on the tradeoff curve of RCs cannot be achieved by exact repair. The exact repair rate region of the simple case of $(n,k,d,r)=(4,3,3,0)$ was characterized in \cite{tian2014characterizing} and it was shown that indeed there is a gap between the optimal tradeoff of functional repair and exact repair. Specifically, functional repair is strictly more powerful than exact repair and its benefits should not be overlooked. 

\par The fundamental finding in \cite{tian2014characterizing} is proven by a computer-aided proof. It turns out that if we focus on a different $(n,k,d,r)$ value other than $(4,3,3,0)$, we can easily prove the same statement ``functional repair strictly outperforms exact repair'' without resorting to the relatively-involved computer-aided-proof approach.

\begin{proposition}\label{prop:exact_vs_functional}
For $(n,k,d,r)=(5,3,2,1)$ and $(5,4,2,1)$, there exists at least one pair of $(\alpha,\beta)$ values such that LRRCs with functional repair can protect a file of size strictly larger than that of LRRCs constrained to exact repair. Furthermore, for these two $(n,k,d,r)$ values, the superiority of functional repair over exact repair occurs in both the MSR and MBR points, unlike the case of $(n,k,d,r)=(4,3,3,0)$ where the superiority occurs only in the interior points.  
\end{proposition}

\begin{IEEEproof}[Proof of Proposition~\ref{prop:exact_vs_functional}]
The proof is by contradiction. Consider $(n,k,d,r)=(5,3,2,1)$ or $(5,4,2,1)$. The following arguments work for both cases. By Propositions~\ref{prop:result_SHS} and \ref{prop:result_DHS}, we know that the tradeoff of the best SHS scheme is the same as that of BHS and that DHS strictly outperforms BHS for at least one pair of $(\alpha,\beta)$ values. Suppose now that there exist exact repair LRRCs with DHS that can achieve the entire optimal tradeoff in \eqref{eq:new-tradeoff}. Since such a code is an exact repair code, the same helper nodes that can repair a failed node at time $\tau=1$ can be used to repair the same failed node at any other time $\tau$. Specifically, instead of having $D_{\tau}(\{F_i\}_{i=1}^{\tau}, \{U_j\}_{j=1}^{\tau})$ that changes over time, we can simply set 
\begin{align}
D_{\tau}(\{F_i\}_{i=1}^{\tau}, \{U_j\}_{j=1}^{\tau})=D_1(F_{\tau}, U_{\tau}).\nonumber
\end{align}
The reason is that in exact repair the packets on the nodes are the same at any time, so we can reuse the helper choice in time 1 and the resulting new code should still meet the reliability requirement. Therefore, the considered exact repair LRRC with DHS can be converted to an exact repair LRRC with SHS with the same tradeoff curve \eqref{eq:new-tradeoff}. This, however, yields a contradiction with Proposition~\ref{prop:result_SHS} that states that with SHS we cannot protect a file of size larger that that in \eqref{eq:BHS-tradeoff}.

\par If we compare the tradeoff curve \eqref{eq:new-tradeoff} of functional repair and the tradeoff curve \eqref{eq:BHS-tradeoff} of the best possible exact repair, see Fig.~\ref{fig:storage_vs_bandwidth_(5-3-2)}, it is clear that the superiority of functional repair over exact repair occurs in both the MSR and MBR points. Hence, the proof is complete.
\end{IEEEproof}

\par We can see that the proof of Proposition~\ref{prop:exact_vs_functional} provides a new simple proof technique that can show that exact repair cannot achieve the performance of functional repair under LRRCs by designing a DHS scheme that strictly outperforms all SHS schemes.

\section{When Can DHS/SHS Outperform BHS?}\label{sec:mincut}

In this section, we prove Proposition~\ref{prop:result_conditions}, which focuses on answering the question: Given $(n,k,d,r)$ values, whether there exists a DHS/SHS scheme that outperforms the baseline BHS scheme. 
\subsection{The $(n,k,d,r)$ Values For Which BHS is Optimal}
For easier reference, we reproduce the converse part of Proposition~\ref{prop:result_conditions} as the following proposition. 
\begin{proposition} \label{prop:neg_gen}
If $k\leq \left\lceil\frac{n-r}{n-d-r}\right\rceil$, then for any arbitrary DHS scheme $A$, we have
\begin{align} \label{eq:neg_gen}
\min_{G\in \mathcal{G}_A}\min_{t\in \DC(G)}\mincut_G(s,t)=\sum_{i=0}^{k-1}\min ((d-i)^+\beta,\alpha).
\end{align}
Specifically, even the most intelligent helper selection will have the same tradeoff curve \eqref{eq:Dimakis} as BHS.
\end{proposition}

Before presenting the proof of Proposition~\ref{prop:neg_gen}, we introduce the following definition and lemma.
\begin{definition}\label{def:m-set} A set of $m$ active storage nodes (input-output pairs) of an IFG is called an $m$-set if the following conditions are satisfied simultaneously. (i) Each of the $m$ active nodes has been repaired at least once; and (ii) Jointly the $m$ nodes satisfy the following property: Consider any two distinct active nodes $x$ and $y$ in the $m$-set and without loss of generality assume that $x$ was repaired before $y$. Then there exists an edge in the IFG that connects $x_\text{out}$ and $y_\text{in}$.
\end{definition}
\par In some way (if grouping the two nodes $u_\text{in}$ and $u_\text{out}$ together as a single node), an $m$-set can be viewed as the generalization of the $m$-{\em{clique}} for the IFGs. Notice that a triangle in a graph is also a 3-set according to the above definition. Now, we start the proof by stating the following lemma which is the core of the proof.

\begin{lemma} \label{clm:set} Fix the helper selection scheme $A$. There exists an IFG $G\in \mathcal{G}_A(n,k,d,r,\alpha,\beta)$ satisfying that at least one $\left\lceil \frac{n-r}{n-d-r}\right\rceil$-set exists in its set of active nodes.
\end{lemma}

\begin{IEEEproof}[Proof of Lemma~\ref{clm:set}]
We prove this lemma by explicit construction. Start first with a graph $G^{\text{init}}$ such that all its active nodes have been failed/repaired before. Define $V$ as the set of active nodes of $G^{\text{init}}$ corresponding to physical storage nodes $\{1,2,\dots,r\}$ where $r$ is the system parameter that limits the maximum number of temporarily unavailable nodes. Now, fail and repair the nodes $\{r+1,r+2,\dots,n\}$ in this order with $V$ being the set of unavailable helper nodes fixed for all times of repair, i.e., we fail node $(r+1)$ first and we repair it under the fact that the nodes in set $V$ are all unavailable, then we fail node $(r+2)$ and repair it with $V$ being unavailable too and so on. The final IFG we get is denoted as graph $G$.

\par We prove that $G$ has at least one $\left\lceil \frac{n-r}{n-d-r}\right\rceil$-set by proving the following stronger claim: Consider any integer value $m\geq 1$. Denote the set of active nodes of $G \backslash V$ by $V^c$. There exists an $m$-set in every subset of $(m-1)(n-d-r)+1$ active nodes of $V^c$. 

\par We first describe how to use the above claim and then provide the corresponding proof. Since the $G$ we consider has $n$ active nodes in total, then $|V^c|=n-r$. By solving the largest $m$ value satisfying  $(m-1)(n-d-r)+1\leq |V^c|=n-r$, the above stronger claim implies that $V^c$ must contain a $\left\lceil \frac{n-r}{n-d-r}\right\rceil$-set. Lemma~\ref{clm:set} is thus proven.

\par We now prove this claim by induction on the value of $m$. When $m=1$, by the definition of the $m$-set, any group of 1 active node in $V^c$ forms a 1-set. The claim thus holds naturally.

\par Suppose the claim is true for all $m<m_0$, we now claim that in every group of $(m_0-1)(n-d-r)+1$ active nodes of $V^c$ there exists an $m_0$-set. The reason is as follows. Given an arbitrary, but fixed group of $(m_0-1)(n-d-r)+1$ active nodes of $V^c$, we use $y$ to denote the youngest active node in this group (the one which was repaired last). Obviously, there are $(m_0-1)(n-d-r)$ active nodes in this group other than $y$. On the other hand, since any newcomer accesses $d$ helpers out of the surviving nodes, during its repair, node $y$, when it was repaired, was able to avoid connecting to at most $(n-r-1)-d$ surviving nodes of $V^c$. Therefore, out of the remaining $(m_0-1)(n-d-r)$ active nodes in this group, node $y$ must be connected to at least $((m_0-1)(n-d-r))-(n-r-1-d)=(m_0-2)(n-d-r)+1$ of them. By induction, among those $\geq (m_0-2)(n-d-r)+1$ nodes of $V^c$, there exists an $(m_0-1)$-set. Since, by our construction, $y$ is connected to {\em all} nodes in this $(m_0-1)$-set, node $y$ and this $(m_0-1)$-set jointly form an $m_0$-set. The proof of the stronger argument and hence Lemma~\ref{clm:set} is thus complete.
\end{IEEEproof}
\begin{IEEEproof}[Proof of Proposition~\ref{prop:neg_gen}] We now prove \eqref{eq:neg_gen}. Consider an IFG $G\in \mathcal{G}_A$ that satisfies Lemma~\ref{clm:set}. Since $k\leq \left\lceil \frac{n-r}{n-d-r} \right\rceil$ we can construct a data collector of $G$ that connects to $k$ nodes out of the nodes of the $\left\lceil \frac{n-r}{n-d-r} \right\rceil$-set in $G$. Call this data collector $t_0$. If we focus on the edge cut that directly separates source $s$ and the $k$ node pairs connected to $t_0$, one can use the same analysis as in \cite[Lemma 2]{dimakis2010network} and derive ``$\mincut(s,t_0)\leq \sum_{i=0}^{k-1}\min ((d-i)^+\beta,\alpha)$'' for the given $G\in \mathcal{G}_ A$ and the specific choice of $t_0$. Therefore, we have
\begin{align}\label{eq:worse_br}
\min_{G\in\mathcal{G}_A}\min_{t\in \DC(G) } \mincut_G(s,t)\leq \sum_{i=0}^{k-1}\min((d-i)^+\beta,\alpha).
\end{align}
On the other hand, by definition we have
\begin{align}
\min_{G\in\mathcal{G}_A}\min_{t\in \DC(G) } \mincut_G(s,t)\geq \min_{G\in\mathcal{G}}\min_{t\in \DC(G) } \mincut_G(s,t). \label{eq:new2_better}
\end{align}
Then by \eqref{eq:worse_br}, \eqref{eq:new2_better}, and \eqref{eq:ex_low_b}, we have proven that whenever $k\leq \left\lceil \frac{n-r}{n-d-r}\right\rceil$, equality \eqref{eq:neg_gen} is true. Hence, the proof is complete.
\end{IEEEproof}

\subsection{The Achievability Proof: Description of a New Helper Selection Scheme} \label{sec:MFHS}
\begin{figure*}
\centering
\includegraphics[width=7.1in,keepaspectratio]{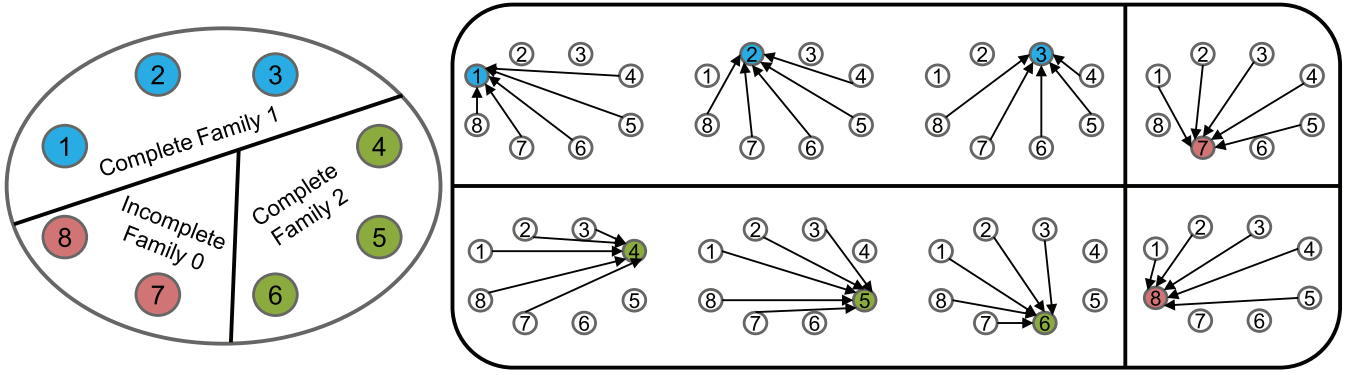}
\caption{The MFHS scheme for $(n,d,r)=(8,4,1)$ and the illustration of the repair process of each of the 8 nodes. Each newcomer may choose to access $(d+r)=5$ helpers, as illustrated in the arrows. However, only $d$ of them will be actually accessed since we assume $r=1$ of the helpers may be temporarily unavailable.}
\label{fig:mfhs}
\end{figure*}

\par For easier reference, we reproduce the achievability part of Proposition~\ref{prop:result_conditions} as the following proposition. 
\begin{proposition} \label{prop:reproduced-ach}
If $\min(d+1,k)> \left\lceil\frac{n}{n-d-r}\right\rceil$, then there exists an SHS scheme and a pair of $(\alpha,\beta)$ such that 
\begin{align} \label{eq:ccw-ach-new}
\min_{G\in \mathcal{G}_A}\min_{t\in \DC(G)}\mincut_G(s,t)>\sum_{i=0}^{k-1}\min ((d-i)^+\beta,\alpha).
\end{align}
\end{proposition}

We prove the above result by explicit construction. In this subsection we will first describe how we choose the helpers and we will analyze its performance in the next subsection. The proposed scheme is called \emph{modified family helper selection (MFHS)} scheme, which is based on the family helper selection (FHS) scheme in \cite{ahmad2014when,arxiv1,arxiv2} that was originally devised for the case of $r=0$, i.e., all helper nodes are always available. 

\par The MFHS can be described as follows. First, we arbitrarily sort all storage nodes and denote them by $1$ to $n$. Then, we define a {\em complete family} as a group of $(n-d-r)$ physical nodes. The first $(n-d-r)$ nodes are grouped as the first complete family, the second $(n-d-r)$ nodes are grouped as the second complete family and so on. In total, we have $\left\lfloor \frac{n}{n-d-r}\right\rfloor$ complete families. The remaining $n\bmod(n-d-r)$ nodes, if there is any, are grouped as an {\em incomplete family}. For any node $F$, if $F$ belongs to a complete family, we use $\overline{D}(F)$ to denote the set of nodes outside the family of $F$. Since the family of node $F$ has $(n-d-r)$ nodes, $\overline{D}(F)$ contains exactly $n-(n-d-r)=d+r$ nodes. If $F$ belongs to an incomplete family, we use $\overline{D}(F)$ to denote the set of nodes from the first node to the $(d+r)$-th node (recall we sorted the nodes in the very beginning). Again $\overline{D}(F)$ contains exactly $(d+r)$ nodes.    

\par One can view the set $\overline{D}(F)$ as the {\em candidate} helper set when node $F$ fails. Specifically, when node $F$ fails and nodes in $U$, $|U|\leq r$, are unavailable, we choose the helper set of node $F$ by $D(F,U)=\overline{D}(F)\backslash U$. Note that we will have $|\overline{D}(F)\backslash U|=(d+r)-r$ if $|U|=r$ and $U\subseteq \overline{D}(F)$. If the unavailable node set is smaller $|U|<r$ or if the unavailable node set is not all within $\overline{D}(F)$, then we simply let node $F$ access the first $d$ available nodes (those with the smallest node indices) in $\overline{D}(F)$ for repair.

\par For example, suppose that $(n,d,r)=(8,4,1)$. There are $2$ complete families, $\{1,2,3\}$ and $\{4,5,6\}$, and $1$ incomplete family, $\{7,8\}$. See Fig.~\ref{fig:mfhs} for illustration. Then suppose node $4$ fails. Since node 4 belongs to a complete family $\{4,5,6\}$, $\overline{D}(4)=\{1,2,3,7,8\}$ since nodes 1, 2, 3, 7, and 8 are outside the family of node 4. Therefore, if node 2 is temporarily unavailable $U=\{2\}$, the newcomer will then access nodes $\overline{D}(4)\backslash U=\{1,3,7,8\}$ for repair. If it is node 8 being unavailable, then the newcomer will access $\overline{D}(4)\backslash\{8\} =\{1,2,3,7\}$ for help. Similarly, if node $7$ fails, then since node 7 belongs to an incomplete family $\{7,8\}$, the corresponding candidate helper set contains the first $(d+r)=5$ nodes $\overline{D}(7)=\{1,2,3,4,5\}$. If node $2$ is unavailable ($U=\{2\}$), then the helpers become $\overline{D}(7)\backslash U=\{1,3,4,5\}$. If, say node 8 is unavailable ($U=\{8\}$), the set $\overline{D}(7)\backslash U=\{1,2,3,4,5\}$ now contains $5>d=4$ nodes. In this scenario, we simply let the newcomer access $\{1,2,3,4\}$, the first $d=4$ nodes of $\overline{D}(7)$, for repair.

\subsection{The Achievability Proof: Analysis of The Modified Family Helper Selection} \label{subsec:mfhs}
In the following, we analyze the performance of the modified family helper selection scheme  (MFHS). Before we analyze the performance, we introduce some useful definitions. 

\par {\em Definition of the family index vector:} Notice that the MFHS scheme has in total $\left\lceil\frac{n}{n-d-r}\right\rceil$ families, which we index from $1$ to $\left\lceil\frac{n}{n-d-r}\right\rceil$. However, since the incomplete family has different properties from the complete families, we index the incomplete family by the family index $0$. The family indices thus become from $1$ to $c\stackrel{\Delta}{=}\left\lfloor\frac{n}{n-d-r}\right\rfloor$ and then $0$, where $c$ is the index of the last Complete family. If there is no incomplete family, we omit the index $0$. Moreover, notice that any member of the incomplete family has $\overline{D}(F)=\{1,\cdots, d+r\}$. That is, for an incomplete family node $F$, $\overline{D}(F)$ contains {\em all} the members of the first $(c-1)$ complete families and only the first $(d+r)-(n-d-r) (c-1)=n\bmod(n-d-r)$ members of the last complete family $c$. Among the $(n-d-r)$ members in the last complete family, {\em we add a negative sign to the family indices of those who will ``not'' be helpers for the incomplete family.}

\par We use the notation $FI(n_0)$ to denote the family index of node $n_0$. We can now list the family indices of the $n$ nodes as an $n$-dimensional {\em family index vector} defined as $(FI(1),FI(2),
\dots,FI(n))$. Considering the same example above where $(n,d,r)=(8,4,1)$, the {\em family index vector} is $(1,1,1,2,2,-2, 0,0)$.

\begin{figure}[h!]
\centering
\includegraphics[width=0.475\textwidth]{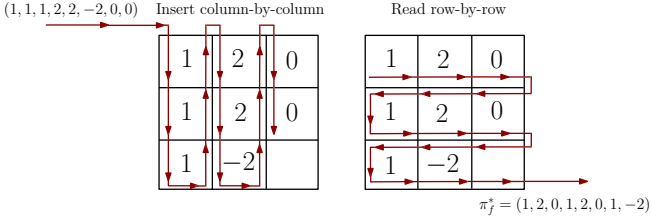}
\caption{The construction of the RFIP for $(n,d,r)=(8,4,1)$.}
\label{fig:rfip}
\end{figure}

\par {\em Definitions of the family index permutation and RFIP:} A {\em family index permutation} is a permutation of the family index vector, which we denote by $\pi_f$. Using the previous example, one instance of family index permutations is $\pi_f=(1,1,0,2,0,-2,1,2)$. A rotating family index permutation (RFIP) $\pi_f^*$ is a special family index permutation that puts the family indices of the family index vector in an $(n-d-r)\times \left\lceil \frac{n}{n-d-r}\right\rceil$ table column-by-column and then reads it row-by-row. Fig.~\ref{fig:rfip} illustrates the construction of the RFIP for the case of $(n,d,r)=(8,4,1)$. The input is the family index vector $(1,1,1,2,2,-2,0,0)$ and the output RFIP $\pi_f^*$ is $(1,2,0, 1,2,0,1,-2)$.

\par We now analyze the performance of the MFHS scheme.

\begin{proposition} \label{prop:low_b}
Consider any given MFHS scheme $F$ with the corresponding IFGs denoted by $\mathcal{G}_F(n,k,d,r,\alpha,\beta)$. We have that
\begin{align}
\min_{G\in\mathcal{G}_F} \min_{t\in\DC(G)}&\mincut_G(s,t) = \nonumber\\
&\min_{\forall \pi_f} \sum_{i=1}^{k}\min ((d-y_i(\pi_f))^+\beta,\alpha),\label{eq:low_b}
\end{align}
where $\pi_f$ can be any family index permutation and $y_i(\pi_f)$ is computed as follows. If the $i$-th coordinate of $\pi_f$ is $0$, then $y_i(\pi_f)$ returns the number of $j$ satisfying both (i) $j<i$ and (ii) the $j$-th coordinate $>0$. If the $i$-th coordinate of $\pi_f$ is not $0$, then $y_i(\pi_f)$ returns the number of $j$ satisfying both (i) $j<i$ and (ii) the absolute value of the $j$-th coordinate of $\pi_f$ and the absolute value of the $i$-th coordinate of $\pi_f$ are not equal. For example, if $\pi_f=(1,2,-2,1,1,0,0,1,2,-2)$, then $y_6(\pi_f)= 4$ and $y_{10}(\pi_f)= 6$.
\end{proposition}

The proof of this proposition will be provided in the end of this subsection. 

\par We notice that computing the right-hand side of \eqref{eq:low_b} requires searching over all possible permutations $\pi_f$. The following proposition shows that when focusing on the minimum bandwidth repair (MBR) point, one can further simplify the expression.
\begin{proposition} \label{prop:mbr}
Consider any $(n,k,d,r)$ values and the MFHS scheme. The MBR point of the MFHS scheme is
\begin{align} \label{eq:gamma}
\alpha_{\mbr}=d\beta_{\mbr}=\frac{d\mathcal{M}}{\sum_{i=1}^{k} (d-y_i(\pi_f^*))^+}
\end{align}
where $\pi_f^*$ is the rotating family index permutation (RFIP). 
\end{proposition}
\par The proof of Proposition~\ref{prop:mbr} is relegated to Appendix~\ref{app:mbr_proof}. 

Proposition~\ref{prop:mbr} directly implies Proposition~\ref{prop:reproduced-ach}, the achievability part of Proposition~\ref{prop:result_conditions}. The reason is as follows. We first notice that by the definition of $y_i(\cdot)$, we always have $y_i(\pi^*_f)\leq i-1$. Suppose $\min(d+1,k)>\left\lceil\frac{n}{n-d-r}\right\rceil$ and consider the MFHS scheme. Since there are exactly $\left\lceil\frac{n}{n-d-r}\right\rceil$ number of families (including both complete and incomplete families), among the first $\min(d+1,k)$ indices of a family index permutation $\pi_f^*$ there is at least one family index that is repeated. Say the $j_1$-th and the $j_2$-th coordinates of $\pi_f^*$ are of the same value where $j_1,j_2\leq \min(d+1,k)$. Without loss of generality, we assume $j_1<j_2$. Then, by the definition of $y_i(\cdot)$, we have $y_{j_2}(\pi^*_f)< j_2-1$ with a strict inequality since the $j_1$-th coordinate of $\pi_f^*$ will not contribute to $y_{j_2}(\pi^*_f)$. Letting $q=\min(d+1,k)$, we thus have
\begin{align}
\sum_{i=1}^k (d-y_i(\pi_f^*))^+&=\sum_{i=1}^{q}(d-y_i(\pi_f^*))+\sum_{i=q+1}^k(d-y_i(\pi_f^*))^+\label{eq:rev-new-mbr-ccw2}\\
&> \sum_{i=1}^{q}(d-(i-1))+\sum_{i=q+1}^k(d-(i-1))^+ \label{eq:rev-new-mbr-ccw1}\\
&=\sum_{i=1}^k (d-(i-1))^+. \label{eq:rev-new-mbr}
\end{align}
where \eqref{eq:rev-new-mbr-ccw2} follows from $y_i(\pi_f^*)\leq (i-1)$ so that we can remove the $()^+$ when $i=1$ to $\min(d+1,k)$ without changing the value;  \eqref{eq:rev-new-mbr-ccw1} follows from that $y_{j_2}(\pi^*_f)< (j_2-1)$ and that $y_i(\pi_f^*)\leq (i-1)$ for arbitrary $i$; and \eqref{eq:rev-new-mbr} follows from $(d-(i-1))\geq 0$ for all $i=1$ to $\min(d+1,k)$. 

\par We now compare the MBR points of the MFHS and the BHS scheme. The MBR point of the MFHS scheme is described by \eqref{eq:gamma} while the MBR point of the BHS scheme is described by 
\begin{align}
\alpha_{\mbr}=d\beta_{\mbr}=\frac{d\mathcal{M}}{\sum_{i=1}^{k} (d-(i-1))^+}. \label{eq:28}
\end{align}
Ineq.~\eqref{eq:rev-new-mbr} then implies that the MFHS strictly outperforms BHS by having strictly smaller storage/BW since \eqref{eq:gamma} is strictly less than \eqref{eq:28}.

\par It is worth mentioning that the result in Proposition~\ref{prop:low_b} is weaker than the results in Section~\ref{subsec:DHS-better} in the following sense. The storage-bandwidth tradeoff curve in Proposition~\ref{prop:low_b} is based purely on a min-cut analysis similar to those in \cite{dimakis2010network}, while relying on the assumption that random linear network coding (RLNC) with sufficiently large finite fields can attain the min-cut capacity for infinitely many IFGs. Also see the discussion in \cite{wu2010existence}. In contrast, the code existence result in Section~\ref{subsec:DHS-better} is in the strongest sense since we provide explicit binary code construction and then directly analyze its performance, see Proposition~\ref{prop:absolute-optimal}, without using any min-cut analysis.

\par For some class of $(n,k,d,r)$ combinations, it is possible to derive explicit code constructions without relying on the RLNC-based assumption. The code constructions are rather involved and for that reason we omit them from this work and provide them in \cite{arxiv_codes}.

We close this subsection by providing the proof of Proposition~\ref{prop:low_b}.

\begin{IEEEproof}[Proof of Proposition~\ref{prop:low_b}] Recall that the MFHS scheme specifies the helper candidate set $\overline{D}(i)$ for nodes $i=1$ to $n$ based on the concepts of complete and incomplete families. In the following discussion, we assume that the helper candidate set $\overline{D}(i)$ is generated by the given MFHS scheme. 

\par Using the same proof technique of \cite[Proposition~5]{arxiv1} and \cite[Lemma~2]{dimakis2010network}, we can get the following lower bound on the smallest possible mincut of an IFG generated by $A$
\begin{align}
\min_{G\in \mathcal{G}_A}\min_{t\in \DC(G)}&\mincut(s,t)\geq\nonumber\\
&\min_{\mathbf{p}\in \mathcal{P}}\sum_{i=1}^{k}\min ((d-z_i(\mathbf{p}))^+\beta,\alpha) \label{eq:low_b_gen},
\end{align}
where $\mathbf{p}$ is a $k$-dimensional integer-valued vector, $\mathcal{P}=\{(p_1,p_2,\cdots,p_k):\forall i \in\{1,\cdots,k\}, 1\leq p_i\leq n\}$ and $z_i(\mathbf{p})=|\{p_j:j<i, p_j\in \overline{D}(p_i)\}|$. For example, suppose $(n,k,d,r)=(6,4,2,1)$, $\overline{D}(3)=\{1,4,5\}$, and $\mathbf{p}=(1,2,1,3)$. Since $p_4=3$, we have $z_4(\mathbf{p})=|\{p_j: j<4, p_j\in \overline{D}(3)\}|= 1$. (The double appearances of $p_1=p_3=1$ are only counted as one.)

\par The main intuition behind \eqref{eq:low_b_gen}, is that for any source $s$ and data collector $t$, we consider the min-cut $(V,V^c)$ separating $t$ from $s$. That is, $V$ and $V^c$ form a partition of the nodes in the IFG; $s\in V$ and $t\in V^c$; and the edges from $V$ to $V^c$  is the minimum edge cut. Since $t\in V^c$, set $V^c$ contains at least $k$ intermediate nodes (the nodes $\neq t$). Denote the $k$ oldest intermediate nodes in $V^c$ by $u_1$ to $u_k$. We denote the node index of each intermediate node $u_i$ by $NI(u_i)$. Note that some $u_i$ and $u_j$ may have the same index $NI(u_i)=NI(u_j)$ since $u_i$ and $u_j$ are intermediate nodes in the IFG, not the actual physical nodes. We then choose the $\mathbf{p}$ vector  by $\mathbf{p}=(NI(u_1),NI(u_2),\cdots, NI(u_k))$.

\par With the above construction, if we examine the definition of $z_i(\mathbf{p})$ in \eqref{eq:low_b_gen}, we can easily see that the function $z_i(\mathbf{p})$ returns an {\em upper bound} of the number of edges entering $u_{i,\text{in}}$ in the IFG from some $u_{j,\text{out}}$ satisfying $j<i$. Therefore, $(d-z_i(\mathbf{p}))^+$  represents a {\em lower bound} of the number of edges entering $u_{i,\text{in}}$  that are {\em not} from $u_{j,\text{out}}$ with $j<i$. Therefore, each $u_i$ will contribute at least $\min ((d-z_i(\mathbf{p}))^+\beta,\alpha)$ to the min-cut value. By summing over  all $u_i$, we have \eqref{eq:low_b_gen}. Since the analysis is quite standard, see \cite[Lemma~2]{dimakis2010network}, we omit the detailed proof of \eqref{eq:low_b_gen}.

\par Next, we will prove that
\begin{align}\label{eq:upper_bound}
\min_{G\in\mathcal{G}_F} \min_{t\in\DC(G)}&\mincut_G(s,t) \leq \nonumber\\
&\min_{\forall \pi_f} \sum_{i=1}^{k}\min (\left(d-y_i(\pi_f)\right)^+\beta,\alpha).
\end{align}
\par The reason is the following. Denote the smallest IFG in $\gf(n,k,d,r,\alpha,\beta)$ by $G_0$. Specifically, all its nodes are intact, i.e., none of its nodes has failed before. Denote its active nodes arbitrarily by $1,2,\cdots,n$. Consider the family index permutation of the MFHS scheme $F$ that attains the minimal value of the right-hand side of \eqref{eq:upper_bound} and call it $\tilde{\pi}_f$. Find a vector of node indices $\tilde{\mathbf{p}}$ such that (i) $FI(\tilde{p}_i)=\tilde{\pi}_f(i)$ for $i=1$ to $n$ and (ii) $\tilde{p}_i\neq \tilde{p}_j$ if $i\neq j$. This is always possible since the family index permutation $\tilde{\pi}_f$ can be viewed as transcribing some node index vector $\tilde{\mathbf{p}}$ to the corresponding family indices. 

\par After constructing $\tilde{\mathbf{p}}$, we fail each active node in $\{1,2,\cdots,n\}$ of $G_0$ exactly once starting by failing node $\tilde{p}_1$ to node $\tilde{p}_n$. Along this failing process, at each step of repair, say we are now repairing $\tilde{p}_i$, we choose the unavailable nodes $U_i$ as follows. Among the $(d+r)$ nodes in $D(\tilde{p}_i)$, we first sort them according to their locations in the node index vector $\tilde{\mathbf{p}}$. Namely, if both nodes $\tilde{p}_{j_1}$ and $\tilde{p}_{j_2}$ belong to $D(\tilde{p}_i)$, then we say $\tilde{p}_{j_1}$ is ahead of $\tilde{p}_{j_2}$ if $j_1<j_2$. Once we have sorted the $(d+r)$ nodes in  $D(\tilde{p}_i)$, we let the {\em last} $r$ nodes of $D(\tilde{p}_i)$ to be temporarily unavailable during the repair of node $\tilde{p}_i$. Therefore, the helpers of the newcomer $\tilde{p}_i$ must be the first $d$ nodes of $D(\tilde{p}_i)$. After repairing all $n$ nodes according to the above description, we denote the final IFG as graph $G'$. 

\par We use the following example to demonstrate the above failing/repair process. Let $(n,d,r)=(8,4,1)$ and suppose the minimizing family index permutation is $\tilde{\pi}_f=(1,2,1,-2,0,0,1,2)$. Then, a possible $\tilde{\mathbf{p}}$ is $\tilde{\mathbf{p}}=(1,4,2,6,7,8,3,5)$, which satisfies $(FI(\tilde{p}_1),FI(\tilde{p}_2), \cdots, FI(\tilde{p}_n))=\tilde{\pi}_f$. Using the permutation $\tilde{\mathbf{p}}$, we fail nodes 1, 4, 2, 6, 7, 8, 3, and 5 in this sequence. To illustrate how we choose the unavailable node set $U_i$ when failing node $\tilde{p}_i$, consider the fourth repair operation, for which node 6 fails and we want to repair it. Recall that node 6 belongs to the second complete family $\{4,5,6\}$. Therefore, $\overline{D}(6)=\{1,2,3,7,8\}$. We sort $\overline{D}(6)$ according to their locations in $\tilde{\mathbf{p}}$ and we thus have $\overline{D}(6)=\{1,2,7,8,3\}$. Therefore, we assume $U=\{3\}$ is unavailable and the helper nodes of node 6 are $\{1,2,7,8\}$. Another example is when repairing node 8, i.e., the sixth repair operation. Since node 8 belongs to an incomplete family, the corresponding helper candidate set is $\overline{D}(8)=\{1,2,3,4,5\}$. After sorting, we have $\overline{D}(8)=\{1,4,2,3,5\}$. Therefore, we make node $5$ to be temporarily unavailable when repairing node 8, and the actual helpers of node 8 become $\{1,2,3,4\}$. Note that $\tilde{\mathbf{p}}$ may  not be unique in our construction. For example, $\tilde{\mathbf{p}}=(3, 5, 2, 6, 8, 7, 1, 4)$ is also a possible permutation satisfying $(FI(\tilde{p}_1),FI(\tilde{p}_2), \cdots, FI(\tilde{p}_n))=\tilde{\pi}_f$. Our construction holds for any arbitrary choice of $\tilde{\mathbf{p}}$. 

\par Consider a data collector $t$ in $G'$ that connects to the oldest $k$ newcomers, i.e., nodes $\tilde{p}_1$ to $\tilde{p}_k$. We now analyze the cut-value between the root $s$ and the data collector $t$ using the same arguments as in \cite[Lemma 2]{dimakis2010network}. Consider a special cut $(V, V^c)$ between $t$ and $s$ that are constructed as follows. Initially, we set $V^c=\{t\}$ containing only the data collector. Then, for each $i\in \{1,\dots,k\}$, if $\alpha \leq (d-y_i(\tilde{\pi}_f))^+\beta$ then we add $x_{\out}^{\tilde{p}_i}$ to $V^c$. Namely, the {\em out half} of node $\tilde{p}_i$ is added to $V^c$; Otherwise, we include both $x_{\out}^{\tilde{p}_i}$ and $x_{\inp}^{\tilde{p}_i }$ in $V^c$. With the above construction of $V^c$, it is not hard to see that the cut-value of the cut $(V,V^c)$ is equal to $\sum_{i=1}^{k}\min ((d-y_i(\tilde{\pi}_f))^+\beta,\alpha)$.

\par Since the LHS of \eqref{eq:upper_bound} further takes the minimum over $\mathcal{G}_F$ and all data collectors $t$, we have proven the inequality \eqref{eq:upper_bound}.

\par Thus far, we have that
\begin{align}\label{eq:upper_lower}
\min_{\mathbf{p}\in \mathcal{P}}\sum_{i=1}^{k}\min ((d-z_i(\mathbf{p}))^+\beta,\alpha)\leq& \nonumber\\
\min_{G\in\mathcal{G}_F} \min_{t\in\DC(G)}\mincut_G(s,t)& \leq \nonumber\\
\min_{\forall \pi_f} \sum_{i=1}^{k} \min ((d-&y_i(\pi_f))^+\beta,\alpha).
\end{align}
The remaining step is to prove that
\begin{align}
\min_{\mathbf{p}\in \mathcal{P}}\sum_{i=1}^{k}\min ((d-z_i(\mathbf{p}))^+\beta,\alpha)\geq\nonumber&\\
\min_{\forall \pi_f} \sum_{i=1}^{k}\min ((d-&y_i(\pi_f))^+\beta,\alpha)\label{eq:equality}.
\end{align}
Once we prove \eqref{eq:equality}, we have \eqref{eq:low_b} since \eqref{eq:upper_lower} is true. We prove \eqref{eq:equality} by showing that for any $\mathbf{p}\in \mathcal{P}$ we can find a $\hat{\pi}_f$ such that\footnote{We note that $\hat{\pi}_f$ does not necessarily have to satisfy $\hat{\pi}_f=(FI(p_1), FI(p_2),\cdots, FI(p_n))$, and in fact $\hat{\pi}_f=(FI(p_1), FI(p_2),\cdots, FI(p_n))$ is not always possible. For illustration, consider $\mathbf{p}=(1,1,1,1,\cdots, 1)$, which is a legitimate choice of $\mathbf{p}\in\mathcal{P}$. However, for such $\mathbf{p}$ it is impossible to find a family index permutation satisfying $\hat{\pi}_f=(FI(p_1), FI(p_2),\cdots, FI(p_n))$ since the vector $(FI(p_1), FI(p_2),\cdots, FI(p_n))$ is not a family index permutation.}
\begin{align}
z_i(\mathbf{p})\leq y_i(\hat{\pi}_f),\quad\forall i=1,\cdots, k.\label{eq:new-pi-p}
\end{align}
One can clearly see that the existence of $\hat{\pi}_f$ satisfying \eqref{eq:new-pi-p} for any $\mathbf{p}\in \mathcal{P} $ immediately implies \eqref{eq:equality}. 

\par In our previous work \cite{arxiv1,arxiv2}, we have proven that \eqref{eq:new-pi-p} holds for the case of $r=0$ and arbitrary $(n,k,d)$ values. We will now prove that \eqref{eq:new-pi-p} holds for arbitrary $(n,k,d,r)$ values.

\par A closer look at the definition of $z_i(\cdot)$ in the proof of Proposition~\ref{prop:low_b} shows that different parameter values $(n,k,d,r)$ and $(n',k',d',r')$ can still lead to the same function  $z_i(\cdot)$ for $i=1$ to $n$, provided we have $n=n'$  and $d+r=d'+r'$. The reason is as follows. Suppose we apply the MFHS scheme to two different scenarios $(n,k,d,r)$ and $(n',k',d',r')$. If we have $n=n'$, then the total number of nodes is the same for both scenarios. Since each complete family contains $n-(d+r)$ nodes, if we also have $(d+r)=(d'+r')$, then MFHS will divide the nodes into families in the same way for both scenarios. Since in MFHS, a newcomer requests help from outside its own family, the helper candidate set $\overline{D}(i)$ will again be the same for both scenarios. Since the definition of $z_i(\cdot)$ in the proof of Proposition~\ref{prop:low_b} depends only on the helper candidate set $\overline{D}(j)$, the $z_i(\cdot)$ function will be identical in both scenarios. 

\par We now argue that if two scenarios $(n,k,d,r)$ and $(n',k',d',r')$ satisfying  $n=n'$  and $d+r=d'+r'$, the $y_i(\cdot)$ function in Proposition~\ref{prop:low_b} will again be the same for both scenarios. The reason is as follows. By comparing the definitions of $y_i(\pi_f)$ and $z_i(\mathbf{p})$, one can quickly see that if we choose the family index permutation $\pi_f$ and the $\mathbf{p}$ that satisfy $\pi_f=(FI(p_1),\cdots, FI(p_n))$, then 
\begin{align}
y_i(\pi_f)=z_i(\mathbf{p}), \quad \forall i=1,\cdots, k.\label{eq:marker1}
\end{align} 
Namely, $y_i(\pi_f)$ can be viewed as a {\em transcribed} version of $z_i(\mathbf{p})$ from the node index $\mathbf{p}$ to a family index $\pi_f$ if $\mathbf{p}\in \mathcal{P}$. Since we have shown that $z_i(\cdot)$ will be the same for both scenarios and since when the $(d+r)=(d'+r')$ the node index to family index transcription will be identical for both scenarios, $y_i(\cdot)$ will also be identical for both scenarios. 

\par Consider any arbitrarily given $(n,k,d,r)$ and use it to generate another scenario $(n',k',d',r')$ satisfying $n'=n$, $k'=k$, $d'=d+r$ and $r'=0$. Consider an arbitrarily chosen $\mathbf{p}\in\mathcal{P}$. For the scenario of $(n',k',d',r')$, since $r'=0$, our previous results in \cite{arxiv1,arxiv2} show that there exists a $\hat{\pi}_f$ satisfying \eqref{eq:new-pi-p}. Since the above paragraphs have proven that the functions $z_i(\cdot)$ and $y_i(\cdot)$ are identical  for both scenarios $(n,k,d,r)$ and $(n',k',d',r')$, the $\hat{\pi}_f$ that satisfies \eqref{eq:new-pi-p} for $(n',k',d',r')$ must also satisfy \eqref{eq:new-pi-p} for the given $(n,k,d,r)$ as well. The proof of Proposition~\ref{prop:low_b} is thus complete.
\end{IEEEproof}

\section{Conclusion}\label{sec:conclusion}
We have shown that stationary helper selection (SHS) can be strictly suboptimal by carefully constructing an optimal binary code for $(n,k,d,r)=(5,3,2,1)$ and $(5,4,2,1)$ based on {\em dynamic helper selection} (DHS), where $r$ represents the maximum number of nodes that can be temporarily unavailable. For general $(n,k,d,r)$ values, we have answered the question whether SHS/DHS can outperform blind helper selection (BHS) or not, for a vast majority of $(n,k,d,r)$ values. The results thus provide valuable guidelines for each $(n,k,d,r)$ whether it is beneficial to spend time and design new SHS/DHS schemes or whether one should simply use the basic BHS.

\appendices
\section{The Information Flow Graph} \label{app:ifg}
\par We provide in this appendix the description of the information flow graph (IFG) that was first introduced in \cite{dimakis2010network}. This appendix follows the same IFG description as in \cite{arxiv1}.

\begin{figure}[h!]
\centering
\includegraphics[width=0.45\textwidth]{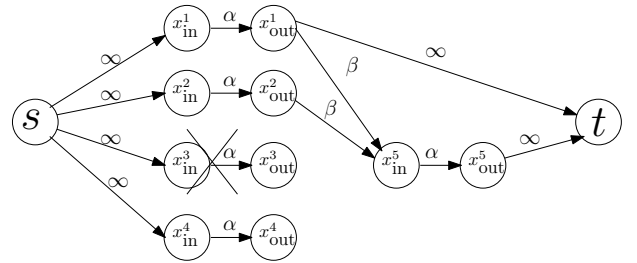}
\caption{An example of the information flow graph with $(n,k,d)=(4,2,2)$.}
\label{fig:ifg}
\end{figure}

\par As shown in Fig~\ref{fig:ifg}, an IFG has three different kinds of nodes. It has a single \emph{source} node $s$ that represents the source of the data object. It also has nodes $x_{\inp}^i$ and $x_{\out}^i$ that represent storage node $i$ of the IFG. A storage node is split into two nodes so that the IFG can represent the storage capacity of the nodes. We often refer to the pair of nodes $x_{\inp}^i$ and $x_{\out}^i$ simply by storage node $i$. In addition to those nodes, the IFG has \emph{data collector} (DC) nodes. Each data collector node is connected to a set of $k$ active storage nodes, which represents the party that is interested in extracting the original data object initially produced by the source $s$. Fig.~\ref{fig:ifg} illustrates one such data collector, denoted by $t$, which connects to $k=2$ storage nodes. 

\par The IFG evolves with time. In the first stage of an information flow graph, the source node $s$ communicates the data object to all the initial nodes of the storage network. We represent this communication by edges of infinite capacity as this stage of the IFG is virtual. See Fig.~\ref{fig:ifg} for illustration. This stage models the encoding of the data object over the storage network. To represent storage capacity, an edge of capacity $\alpha$ connects the input node of storage nodes to the corresponding output node. When a node fails in the storage network, we represent that by a new stage in the IFG where, as shown in Fig.~\ref{fig:ifg}, the newcomer connects to its helpers by edges of capacity $\beta$ resembling the amount of data communicated from each helper. We note that although the failed node still exists in the IFG, it cannot participate in helping future newcomers. Accordingly, we refer to failed nodes by \emph{inactive} nodes and existing nodes by \emph{active} nodes. By the nature of the repair problem, the IFG is always acyclic.

\par Given an IFG $G$, we use $\DC(G)$ to denote the collection of all ${n\choose k}$ {\em data collector nodes} in $G$ \cite{dimakis2010network}. Each data collector $t\in\DC(G)$ represents one unique way of choosing $k$ out of $n$ active nodes when reconstructing the file.

\section{Proof of Proposition~\ref{prop:new2}}\label{app:new2}

\par The statement that BHS is optimal if \eqref{eq:ccw-cond1} holds is a restatement of Proposition~\ref{prop:result_conditions}. We now prove that under the additional assumption $d=r=1$, the BHS scheme is optimal if either $k=3$ or if $k = 4$ and $n \bmod 3\neq 0$.

\par We first give the following definition of an $m$-tree that will be useful in our proof.
\begin{definition} \label{def:tree}
Consider $(n,k,d,r)$ such that $d=1$. Consider an IFG $G$ and a set of $m$ active nodes of $G$ denoted by $x^1,x^2,\dots,x^m$. The set of $m$ active nodes $\{x^1,\dots,x^m\}$ is said to be an $m$-tree if the following two properties hold simultaneously. (a) For $i,j\in \{1,2\cdots, m-1\}$ and $j>i$, $x^i$ is repaired before $x^j$; (b) for any $i=2$ to $m$, there exists a node $b\in\{1,\cdots, i-1\}$ such that $(x^b_\text{out},x^i_\text{in})$ is an edge in $G$.
\end{definition}
The reason that we call the above $m$ nodes an $m$-tree is because since $d=1$, there is exactly 1 edge entering each node $x_\text{in}^i$. The above condition (b) thus implies that each node $x_\text{in}^i$ is {\em connected} to one of the previous nodes $x^1_\text{out}$ to $x^{i-1}_\text{out}$. Therefore, these $m$ nodes form a tree. 

\par We first consider the case of $k=3$, and we state the following claim.

\begin{claim} \label{clm:vertex-cut}
Consider $(n,k,d,r)$ parameters that satisfy that $d=r=1$. For any given DHS scheme $A$ and the corresponding collection of IFGs $\mathcal{G}_A$, we can always find a $G^*\in \mathcal{G}_A$ such that there exists a $3$-tree in its set of active nodes.
\end{claim}

\par We now use the above claim to prove BHS is optimal if $k=3$. Suppose the above claim is true. We let $t^*$ denote the data collector that is connected to the $3$-tree. By properties (a) and (b) in Definition~\ref{def:tree} of a 3-tree, we can see that node $x^1$ is a vertex-cut separating source $s$ and the data collector $t^*$. The min-cut value separating $s$ and $t^*$ thus satisfies $\mincut_{G^*}(s,t^*)=\min(\beta,\alpha)$ for the given  $G^*\in G_A$ and the specific choice of $t^*$. Also note that $\min(\beta,\alpha)=\sum_{i=0}^{k-1}\min((d-i)^+\beta,\alpha)$ since we assume $d=1$ and $k=3$. Combining both, we thus have $\mincut_{G^*}(s,t^*)=\sum_{i=0}^{k-1}\min((d-i)^+\beta,\alpha)$. By the BHS tradeoff curve formula in \eqref{eq:ex_low_b}, we thus have that BHS is optimal when $k=3$ holds.

\begin{IEEEproof}[Proof of Claim~\ref{clm:vertex-cut}]
We prove Claim~\ref{clm:vertex-cut} by explicit construction. Start from any $G\in \mathcal{G}_A$ with all $n$ nodes have been repaired at least once. We choose one arbitrary active node in $G$ and denote it by $w^{(1)}$. We let $w^{(1)}$ fail and denote the newcomer that replaces $w^{(1)}$ by $y^{(1)}$. The helper selection scheme $A$ will choose a helper node (since $d=1$) and we denote that helper node as $x^{(1)}$. The new IFG after this failure and repair process is denoted by $G^{(1)}$. By our construction $x^{(1)}$, as an existing active node, is repaired before the newcomer $y^{(1)}$ and there is an edge $(x^{(1)}_\text{out},y^{(1)}_\text{in})$ in $G^{(1)}$.

Now starting from $G^{(1)}$, we choose another $w^{(2)}$, which is not one of $x^{(1)}$ and $y^{(1)}$ and let this node fail. We use $y^{(2)}$ to denote the newcomer that replaces $w^{(2)}$. The helper selection scheme $A$ will again choose a helper node based on the history of the failure pattern. We denote the new IFG (after the helper selection chosen by scheme $A$) as $G^{(2)}$. If the helper node of $y^{(2)}$ is $x^{(1)}$ or $y^{(1)}$, then the three nodes $(x^{(1)},y^{(1)}, y^{(2)})$ are the $(x^1,x^2,x^3)$ nodes satisfying properties (a) and (b) of Definition~\ref{def:tree} of a $3$-tree. In this case, we can stop our construction and let $G^*=G^{(2)}$ and we say that the construction is complete in the second round. Suppose the above case is not true, i.e., the helper of $y^{(2)}$ is neither $x^{(1)}$ nor $y^{(1)}$. Then, we denote the helper of $y^{(2)}$ by $x^{(2)}$. Note that after this step, $G^{(2)}$ contains two disjoint pairs of active nodes such that there is an edge $(x^{(m)}_\text{out}, y^{(m)}_\text{in})$ in $G^{(2)}$ for $m=1,2$.

\par We can repeat this process for the third time by failing a node $w^{(3)}$ that is none of $\{x^{(m)},y^{(m)}:\forall m=1,2\}$. Again, let $y^{(3)}$ denote the newcomer that replaces $w^{(3)}$ and the scheme $A$ will choose a helper for $y^{(3)}$. The new IFG after this failure and repair process is denoted by $G^{(3)}$. If the helper of $y^{(3)}$ is $x^{(m)}$ or $y^{(m)}$ for some $m=1,2$, then the three nodes $(x^{(m)},y^{(m)}, y^{(3)})$ are the $(x^1,x^2,x^3)$ nodes in Definition~\ref{def:tree} satisfying properties (a) and (b). In this case, we can stop our construction and let $G^*=G^{(3)}$ and we say that the construction is complete in the third round. If the above case is not true, then we denote the helper of $y^{(3)}$ by $x^{(3)}$, and repeat this process for the fourth time and so on.

\par If the construction is not complete in the $m$-th round for some $m\leq \left\lceil \frac{n}{2}\right\rceil-1$, we can always start the $(m+1)$-th round since out of the $n$ nodes, we can always find another node $w^{(m+1)}$ that is none of $\{x^{(m')},y^{(m')}:\forall m'=1,2,\dots, m\}$. Now, suppose that $n$ is odd and the construction is not completed after $m_0=\left\lceil \frac{n}{2}\right\rceil-1$ rounds. In this case, there is only 1 remaining node that is not inside $\{x^{(m)},y^{(m)}:\forall m=1,2,\dots, m_0\}$. Denote that node as $w^{(m_0+1)}$. Fail $w^{(m_0+1)}$ and replace it by $y^{(m_0+1)}$. Since $\{x^{(m)},y^{(m)}:\forall m=1,2,\dots, m_0\}$ and $y^{(m_0+1)}$ cover all $n$ nodes, the helper node of $y^{(m_0+1)}$ must be one of the nodes in $\{x^{(m)},y^{(m)}:\forall m=1,2,\dots, m_0\}$. If the helper node of $y^{(m_0+1)}$ is $x^{(m')}$ or $y^{(m')}$ for some $m'=1,2,\dots, m_0$, then the three nodes $(x^{(m')},y^{(m')}, y^{(m_0+1)})$ form a 3-tree satisfying properties (a) and (b) of Definition~\ref{def:tree}. 

\par For the case when $n$ is even and and the construction is not completed after $m_0=\left\lceil \frac{n}{2}\right\rceil-1$ rounds. In this case, there are 2 remaining nodes that are not inside $\{x^{(m)},y^{(m)}:\forall m=1,2,\dots, m_0\}$. Choose arbitrarily one of them and denote that node as $w^{(m_0+1)}$. Fail $w^{(m_0+1)}$ and replace it by $y^{(m_0+1)}$ while having the other remaining node (the one that is not $w^{(m_0+1)}$) temporarily unavailable when repairing $y^{(m_0+1)}$. Therefore, we have forced $y^{(m_0+1)}$ to connect to an $x^{(m')}$ or $y^{(m')}$ node for some $m'=1,2,\dots,m_0$. Similar to the case for which $n$ is odd, the three nodes $(x^{(m')},y^{(m')}, y^{(m_0+1)})$ form a 3-tree. The proof of Claim~\ref{clm:vertex-cut} is complete.

\end{IEEEproof}

\par Now, we turn our attention to the case when $k=4$ and $n\bmod 3\neq 0$. Similarly, we state the following claim.

\begin{claim} \label{clm:vertex-cut-2}
Consider $(n,k,d,r)$ parameters that satisfy that $d=r=1$ and $n\bmod(3)\neq 0$. For any given DHS scheme $A$ and the corresponding collection of IFGs $\mathcal{G}_A$, we can always find a $G^{**}\in \mathcal{G}_A$ such that there exists a $4$-tree in its set of active nodes.
\end{claim}

\par We now use the above claim to prove BHS is optimal if $k=4$ and $n\bmod 3\neq 0$. Suppose the above Claim~\ref{clm:vertex-cut-2} is true. As we did above, we let $t^{**}$ denote the data collector that is connected to the $4$-tree. By properties (a) and (b) of the definition of a $4$-tree we can again see that node $x^1$ is a vertex-cut separating source $s$ and the data collector $t^{**}$. The min-cut value separating $s$ and $t^{**}$ thus satisfies $\mincut_{G^{**}}(s,t^{**})\leq\min(d\beta,\alpha)=\sum_{i=0}^{k-1}\min((d-i)^+\beta,\alpha)$ for $G^{**}\in G_A$ and the specific choice of $t^{**}$, where the inequality, as discussed before, follows from $x^1$ being a vertex-cut separating $s$ and $t^{**}$ and the equality follows from the assumption that $d=1$ and $k=4$. We thus have that BHS is optimal when $k=4$ and $n\bmod 3\neq 0$.

\begin{IEEEproof} [Proof of Claim~\ref{clm:vertex-cut-2}]
We prove this claim by explicit construction. The construction contains 2 phases. The goal of Phase 1 is to convert all nodes to be either part of a 2-set or part of a 3-tree. We start from time 1, when no node has ever been repaired. Let $V$ denote a subset of nodes. Initially, set $V=\emptyset$. We arbitrarily choose one node in $\{1,\cdots,n\}\backslash V$, say node $w$. We fail node $w$. The newcomer is denoted by $y$ and we let $x$ denote its helper. After repairing $y$, we add both $\{x,y\}$ to the node set $V$. After updating $V$, we again choose arbitrarily a $w\in\{1,\cdots,n\}\backslash V$, fail it, and replace it by a newcomer $y$ with the corresponding helper being $x$. If the helper $x$ is already in $V$, then we add $y$ to $V$. If $x$ is not in $V$, we add both $\{x,y\}$ to $V$. Repeat the above process until $\{1,\cdots, n\}\backslash V=\emptyset$. 
\par Consider two possibilities. If the resulting IFG contains a 4-tree, then our construction is complete. If not, then we argue that all the nodes in $V$ ($n$ nodes in total when the construction terminates) must either be in a 2-set or in a 3-tree but cannot be in both. We prove this by induction. Suppose $V$ contains only disjoint 2-sets or 3-trees during the construction. Consider our iterative construction, for which we choose a node $w\in\{1,\cdots,n\}\backslash V$ and replace it by a newcomer $y$ with the corresponding helper being $x$. If $x\notin V$ already, after adding $\{x,y\}$ to $V$ the new pair $(x,y)$ will form a 2-set that is disjoint to all the previous nodes in $V$. The induction assumption holds. If $x\in V$ already, we claim that $x$ must be part of a 2-set. The reason why $x$ cannot be part of a 3-tree is that if so, then the 3-tree plus the newcomer $y$ will form a 4-tree and we have already ruled out such a possibility by focusing on the case for which the construction does not lead to any 4-tree. 

\par We are now ready for Phase 2. Recall that after Phase 1 all $n$ nodes have been partitioned to be a collection of disjoint 2-sets or 3-trees. Pick arbitrarily a $2$-set in the active nodes of $G^{(1)}$. This is always possible since $n\bmod 3\neq 0$, which implies that the $n$ nodes cannot all be 3-trees. Denote the chosen 2-set by $(v,w)$. Fail node $w$ and during its repair let $v$ be unavailable. Call the newcomer $w'$. If $w'$ connects to a node that belongs to a 3-tree, then the 3-tree and the newcomer $w'$ form a 4-tree. The construction is thus finished/terminated. 

\par If $w'$ connects to a node that belongs to a 2-set, then the 2-set and $w'$ now form a 3-tree. Namely, we have converted $w$, part of a 2-set, to a new node $w'$ being part of a 3-tree. We then fail node $v$ and replace it by a newcomer $v'$. Similarly, if $v'$ connects to a node that belongs to a 3-tree, then the 3-tree and the newcomer $v'$ form a 4-tree. The construction is finished/terminated. If $v'$ connects a node that belongs to a 2-set, then the 2-set and $v'$ now form a 3-tree. Specifically, we have converted $v$, part of a 2-set, to a new node $v'$ being part of a 3-tree. One can see that the above procedure removes the 2-set $(v,w)$ from the IFG and replaces it by $v'$ and $w'$ that participate in two different 3-trees. 

\par We then iteratively repeat the above process to convert all 2-sets into 3-trees.  This is always possible since $n\bmod 3\neq 0$, which implies that the $n$ nodes cannot all be 3-trees. Nonetheless, we cannot repeat this process indefinitely since each round will remove one 2-set and we only have finitely many 2-sets. This implies that the process must terminate after some finite rounds. Specifically, either $w'$ or $v'$ will be connected to a 3-tree and we will have a 4-tree in the end of this construction. The proof of Claim~\ref{clm:vertex-cut-2} is complete.
\end{IEEEproof}

\par Thus far, we have proven the converse part of Proposition~\ref{prop:new2} that BHS is optimal when conditions (i)-(iii) are satisfied. Now, we notice that, under the assumption of $d=r=1$, the statement ``none of (i)-(iii) holds'' is equivalent to ``at least one of the following conditions holds: (a) $n\bmod 3=0$ and $k=4$, or (b) $k\geq 5$.'' The reason is by the simple observation that when focusing on $d=r=1$, condition (i) is equivalent to ``$k\leq 2$'' since $\left\lceil\frac{n-r}{n-d-r}\right\rceil=2$. Therefore, not satisfying (i) to (iii) is equivalent to satisfying one of conditions (a) and (b).

\par In the following, we will prove that there exists an SHS scheme that can outperform BHS when the $(n,k,d,r)$ parameters satisfy at least one of conditions~(a) or (b). Suppose condition~(a) is satisfied. We prove the existence of such SHS scheme by explicit construction.

\par Our construction is as follows. Since $n\bmod 3=0$, we can divide $n$ nodes into $\frac{n}{3}$ groups of 3 nodes. Suppose the file to be protected has size $\mathcal{M}$. We first divide the file into 2 {\em packets}, each of size $\mathcal{M}/2$. We call each packet the {\em systematic} packet, which is analogous to the concept of systematic bits in error control coding. We then use an $(\frac{n}{3}, 2)$ MDS code to protect the systematic packets by adding $\frac{n}{3}-2$ {\em parity} packets. Finally, each group of 3 nodes is associated with one distinct packet (can be either systematic or parity packets). Each packet is then duplicated 3 times and all 3 nodes in the same group will store an identical copy of the packet of that group. 

\par We argue that such a system can satisfy $(n,k,d,r,\alpha,\beta)$ satisfying $k=4$, $d=r=1$, $\alpha=\beta=\frac{\mathcal{M}}{2}$. The reason is as follows. $\alpha=\frac{\mathcal{M}}{2}$ since each node only stores 1 packet of size $\frac{\mathcal{M}}{2}$. Since $k=4>3$, any $k$ nodes must belong to at least 2 different groups and the nodes in these $\geq 2$ groups must collectively contain $\geq 2$ distinct packets. Because we use an $(\frac{n}{3}, 2)$ MDS code to protect the file, one can reconstruct the original file by accessing any $k=4$ nodes. We now consider the repair operation. Suppose a node fails and we consider the other 2 nodes of the same group. Since $r=1$, at least one of the other two nodes must still be available. The newcomer can thus ask the remaining available node to transfer the packet it stores to the newcomer. Therefore exact-repair can be achieved with $d=r=1$ and $\beta=\frac{\mathcal{M}}{2}$. 

\par We now compare the performance to a BHS scheme for the same $(n,k,d,r)$ parameter value. By \eqref{eq:Dimakis}, the tradeoff curve of BHS when $d=r=1$ and $k=4$ becomes
\begin{align}
\sum_{i=0}^{k-1}\min((d-i)^+\beta,\alpha)=\min(\beta,\alpha)\geq \mathcal{M}\label{eq:temp-BHS}
\end{align} 
One can clearly see that our parameter values $\alpha=\beta=\frac{\mathcal{M}}{2}$ do not satisfy \eqref{eq:temp-BHS}. As a result, the above scheme strictly outperforms the BHS scheme. 

\par Suppose now that condition~(b) is satisfied, i.e., $k\geq5$. Our construction is almost identical to the scheme we described for condition~(a). That is, depending on the values of  $n\bmod 3$, we can either divide the $n$ nodes into $\frac{n}{3}$ group of 3 nodes; or $\frac{n-1}{3}-1$ groups of 3 nodes plus 1 group of 4 nodes; or $\frac{n-2}{3}-2$ groups of 3 nodes plus 2 groups of 4 nodes. Regardless of which case we are in, we again divide the file of size $\mathcal{M}$ into 2 packets, each of size $\frac{\mathcal{M}}{2}$. Then we protect the two packets by an $(\left\lfloor\frac{n}{3}\right\rfloor, 2)$ MDS code. Associate each group with one coded packet and let the nodes of each group store an identical copy of that packet. Since every group has at most 4 nodes, any $k\geq 5$ nodes must belong to at least two different groups. Since any 2 packets can be used to recover the original file, the proposed scheme can reconstruct the original file from any $k$ nodes.  By similar reasons as before, exact-repair can also be achieved with $d=r=1$ and $\beta=\frac{\mathcal{M}}{2}$. We now compare the performance to a BHS scheme for the same $(n,k,d,r)$ parameter value. By \eqref{eq:Dimakis}, the tradeoff curve of BHS when $d=r=1$ and $k=5$ is again \eqref{eq:temp-BHS}. The above scheme with $\alpha=\beta=\frac{\mathcal{M}}{2}$ thus strictly outperforms the BHS scheme. The proof of Proposition~\ref{prop:new2} is hence complete.

\section{The Gap Between \eqref{eq:ccw-cond1} and \eqref{eq:ccw-cond2} When $r=1$ and $d=2$}\label{app:gap}
We want to show in the following that, when $r=1$ and $d=2$, \eqref{eq:ccw-cond1} and \eqref{eq:ccw-cond2} cover all the range of parameters that satisfy \eqref{eq:par_cond} except for the points $(n,k,d,r)=(5,3,2,1)$ and $(5,4,2,1)$. When $r=1$ and $d=2$, the LHS of \eqref{eq:ccw-cond1} becomes $\left\lceil\frac{n-1}{n-3}\right\rceil$ and the LHS of \eqref{eq:ccw-cond2} becomes $\left\lceil\frac{n}{n-3}\right\rceil$. Since $d\leq n-1-r$, we must have $n\geq 4$ when $r=1$ and $d=2$. In the following we analyze the gap between the two conditions ``$k\leq \left\lceil\frac{n-1}{n-3}\right\rceil$'' and ``$\min(d+1,k)> \left\lceil\frac{n}{n-3}\right\rceil$'' for different $n$ values. 

\par For $n=4$, \eqref{eq:ccw-cond1} becomes $k\leq \left\lceil\frac{3}{1}\right\rceil=3$ and \eqref{eq:ccw-cond2} becomes $\min(3,k)>\left\lceil\frac{4}{1}\right\rceil=4$. By \eqref{eq:par_cond}, we must have that $k\leq n-1=3$. Therefore, for the scenarios of $n=4$, $d=2$, and $r=1$, all possible $(n,k,d,r)$ values satisfy \eqref{eq:ccw-cond1} and none of them satisfy \eqref{eq:ccw-cond2}.

\par For $n=5$, \eqref{eq:ccw-cond1} becomes $k\leq \left\lceil\frac{4}{2}\right\rceil=2$. By \eqref{eq:par_cond}, we must have that $k\leq n-1=4$. For $k=1,2$, \eqref{eq:ccw-cond1} is satisfied. On the other hand, $k=3,4$ cannot satisfy \eqref{eq:ccw-cond1}. Since \eqref{eq:ccw-cond2} becomes $\min(3,k)>\left\lceil\frac{5}{2}\right\rceil=3$, no $k$ value can satisfy \eqref{eq:ccw-cond2}. We thus have that points $(n,k,d,r)=(5,4,2,1)$ and $(5,3,2,1)$ satisfy neither \eqref{eq:ccw-cond1} nor \eqref{eq:ccw-cond2}.

\par For $n\geq 6$, we first observe that $1<\frac{n-1}{n-3}<\frac{n}{n-3}\leq 2$ whenever $n\geq 6$. The reason is as follows. The first 2 strict inequalities are straightforward. The last inequality follows from that $\frac{n}{n-3}$ is monotonically decreasing with $n$ and $\frac{6}{6-3}=2$. The above observation thus ensures that when $n\geq 6$, \eqref{eq:ccw-cond1} becomes $k\leq \left\lceil\frac{n-1}{n-3}\right\rceil=2$ and \eqref{eq:ccw-cond2} becomes $\min(3,k)>\left\lceil\frac{n}{n-3}\right\rceil=2$. We can see that there is no gap between these two conditions. Therefore, for $n\geq 6$, all possible parameters satisfy either \eqref{eq:ccw-cond1} or \eqref{eq:ccw-cond2}.

\par We have thus shown that the only points that conditions \eqref{eq:ccw-cond1} and \eqref{eq:ccw-cond2} do not cover are the points $(n,k,d,r)=(5,4,2,1)$ and $(5,3,2,1)$.

\section{Proofs of Lemma~\ref{lem:feasibility_ca} and Proposition~\ref{prop:code-existence}}\label{app:ca_existence}

In this appendix, we prove both Lemma~\ref{lem:feasibility_ca} and Proposition~\ref{prop:code-existence} simultaneously.

The proof is by induction with the following two induction conditions: (A) if node $i$ is the parent of node $j$, then jointly nodes $i$ and $j$ contain 3 linearly independent packets, (B) if neither $i$ is a parent of $j$ nor $j$ is a parent of $i$, then jointly nodes $i$ and $j$ contain 4 linearly independent packets. Let time $\tau=0$ be the stage where the storage network is still intact, i.e., no node has failed before. By checking all ${5\choose 2}$ pairs of nodes, we can easily see that the initial code in Fig.~\ref{fig:initialization} satisfies the induction conditions (A) and (B) at $\tau=0$. In the following, we first show that conditions (A) and (B) guarantee that we can always find packets $(Z_b^*,Z_c^*)$ using the regular repair operations and that the induction conditions (A) and (B) will remain satisfied for the new code.

\par Now, let us assume that the induction conditions (A) and (B) are satisfied until time $\tau=\tau_0-1$ and, using the same notation as above, node $a$ is the failed node with nodes $\{b,c\}$ selected as helpers by the CA scheme. We also use the same notation to denote the two packets in node $i$ by $(Y_1^{(i)},Y_2^{(i)})$. We introduce some vector notation to aid in this proof. We first let $\mathbf{m}$ be a $4\times 1$ column vector holding the 4 packets of the file such that $\mathbf{m}^T=\begin{pmatrix}X_1& X_2&X_3&X_4\end{pmatrix}$. Since the file size is 4, we can express each coded packet by a $4\times 1$ column vector $\mathbf{v}$  over the binary field. Specifically, we denote the vectors of the packets in node $i$ by $(\mathbf{v}_1^{(i)},\mathbf{v}_2^{(i)})$, which means that $Y_1^{(i)}=\mathbf{m}^T\mathbf{v}_1^{(i)}$ and $Y_2^{(i)}=\mathbf{m}^T\mathbf{v}_2^{(i)}$. 

\par With the above vector notation, consider node $b$ and denote the linear span of the vectors $(\mathbf{v}_1^{(b)},\mathbf{v}_2^{(b)})$ of node $b$ by $\mathbf{B}$. Specifically, $\mathbf{B}=\sspan(\{\mathbf{v}_1^{(b)}, \mathbf{v}_2^{(b)}\})=\{\mathbf{0},\mathbf{v}_1^{(b)}, \mathbf{v}_2^{(b)},\mathbf{v}_1^{(b)}+ \mathbf{v}_2^{(b)}\}$, where $\mathbf{0}$ is the zero vector. Similarly, denote by $\mathbf{C}$, $\mathbf{D}$, and $\mathbf{E}$ the span of the vectors of nodes $c$, $d$, and $e$, respectively. In the following, we give an equivalent mathematical presentation of the regular repair operations that choose the $(Z_b^*,Z_c^*)$ packets based on the linear spans $\mathbf{B}$, $\mathbf{C}$, $\mathbf{D}$, and $\mathbf{E}$.

\par \emph{Choosing the $Z_b^*$ packet:} Using the above vector notation, how we choose $Z_b^*$ can be rewritten as follows. If $\rank(\mathbf{C}\oplus\mathbf{D})=4$ where $\oplus$ is the {\em sum-space} operator, then construct set $\mathbf{S_1}=\mathbf{D}$. If $\rank(\mathbf{C}\oplus\mathbf{D})\leq 3$, then construct set $\mathbf{S_1}=\mathbf{C}\oplus \mathbf{D}$. If $\rank(\mathbf{C}\oplus\mathbf{E})=4$ then construct set $\mathbf{S_2}=\mathbf{D}$. If $\rank(\mathbf{C}\oplus\mathbf{E})\leq 3$, then construct set $\mathbf{S_2}=\mathbf{C}\oplus \mathbf{E}$. Then, we choose arbitrarily a vector $\mathbf{v}_b\in \mathbf{B}\backslash (\mathbf{S_1}\cup \mathbf{S_2})$ and then send $Z_b^*=\mathbf{m}^T\mathbf{v}_b$. 

\par We now explain the reason why the above new code construction method is equivalent to the previous description of the repair operations. To that end, we notice that whenever $\mathbf{v}_b\notin \mathbf{S_1}$, then the coded packet $Z_b^*$ will satisfy Condition~1 in our construction. Similarly, whenever $\mathbf{v}_b\notin \mathbf{S_2}$, the coded packet $Z_b^*$ will satisfy Condition~2 in our construction. Since we choose $\mathbf{v}\in B\backslash (\mathbf{S_1}\cup \mathbf{S_2})$, the $Z_b^*$ will simultaneously satisfy both conditions. 

\par We will argue now that we can always find such a vector $\mathbf{v}_b$. To that end, we will first prove that regardless how we construct $\mathbf{S_i}$, $i=1,2$, we always have $\rank(\mathbf{B}\cap \mathbf{S_i})\leq 1$. Since the construction of $\mathbf{S_i}$ are symmetric (one focusing on spaces $\mathbf{C}$ and $\mathbf{D}$, and the other on spaces $\mathbf{C}$ and $\mathbf{E}$), we prove only $\rank(\mathbf{B}\cap \mathbf{S_1})\leq 1$. Consider two cases. Case 1: $\rank(\mathbf{C}\oplus \mathbf{D})=4$. In this case, $\mathbf{S_1}=\mathbf{D}$, then we have
\begin{align}
\rank(\mathbf{B} \cap \mathbf{S_1})&=\rank(\mathbf{B}) + \rank(\mathbf{D}) - \rank(\mathbf{B}\oplus \mathbf{D})\nonumber\\
&\leq 2 + 2 - \rank(\mathbf{B} \oplus \mathbf{D}) \label{eq:rev1}\\
&\leq 1, \label{eq:rev2}
\end{align}
where \eqref{eq:rev1} follows from that the rank of the space of each node is at most 2, and \eqref{eq:rev2} follows from that the induction conditions (A) and (B) imply that the rank of the sum space of two nodes is either 3 or 4, depending on whether one is the parent of the other. 

\par Case 2: $\rank(\mathbf{C}\oplus \mathbf{D})\leq 3$. In this case $\mathbf{S_1}=\mathbf{C}\oplus \mathbf{D}$ and we have
\begin{align}
\rank(\mathbf{B}\cap \mathbf{S_1})&= \rank(\mathbf{B} \cap (\mathbf{C}\oplus \mathbf{D}))\nonumber\\
&=\rank(\mathbf{B}) + \rank(\mathbf{C}\oplus \mathbf{D}) - \nonumber\\
&\quad\quad \rank(\mathbf{B}\oplus (\mathbf{C}\oplus \mathbf{D}))\nonumber\\
&\leq 2 + 3 - \rank(\mathbf{B}\oplus (\mathbf{C}\oplus \mathbf{D})) \label{eq:rev3}\\
&=2+3-4\label{eq:rev4}\\
&= 1, \nonumber
\end{align}
where \eqref{eq:rev3} follows from that $\rank(\mathbf{B})\leq 2$ and $\rank(\mathbf{C}\oplus \mathbf{D})\leq 3$, and \eqref{eq:rev4} follows from the fact that because we use the Clique-Avoiding algorithm, among any the three nodes $b$, $c$, and $d$, at least two of them do not have the parent-child relationship. By the induction assumption (B), we must have $\rank(\mathbf{B}\oplus \mathbf{C}\oplus \mathbf{D})=4$. 

\par The above arguments prove that $\rank(\mathbf{B}\cap \mathbf{S_i})\leq 1$ for $i=1,2$. If we count the number of elements in $\mathbf{B}\cap \mathbf{S_1}$ and $\mathbf{B}\cap \mathbf{S_2}$, then we must have $|\mathbf{B} \cap \mathbf{S_i}|\leq 2^1=2$  for $i=1,2$. Therefore, the size of $((\mathbf{B}\cap \mathbf{S_1})\cup(\mathbf{B}\cap \mathbf{S_2})) $ is at most 3 since both $\mathbf{B}\cap \mathbf{S_1}$ and $\mathbf{B}\cap \mathbf{S_2}$ are linear subspaces and both thus contain the zero vector as a common element. As a result, 
\begin{align}
|\mathbf{B}\backslash (\mathbf{S_1}\cup \mathbf{S_2})|&=|\mathbf{B}|-|((\mathbf{B}\cap \mathbf{S_1})\cup(\mathbf{B}\cap \mathbf{S_2}))|\nonumber\\
&\geq 2^2-3\nonumber\\
&=1.\nonumber
\end{align}
Therefore, there exists at least one vector $\mathbf{v}_b\in \mathbf{B}\backslash (\mathbf{S_1}\cup \mathbf{S_2})$.

\par \emph{Choosing the $Z_c^*$ packet:} Using the above notation, how we choose $Z_c^*$ can be rewritten as follows. Recall that $\mathbf{v}_b$ is the vector for the coded packet $Z_b^*$. We argue that the construction of $Z_c^*$ in Section~\ref{subsec:DHS-better} is equivalent to the following construction. That is, we choose arbitrarily a vector $\mathbf{v}_c \in \mathbf{C}\backslash (\mathbf{v}_b\oplus \mathbf{D}))\cup(\mathbf{v}_b\oplus \mathbf{E}))$ and then send $Z_c^*=\mathbf{m}^T\mathbf{v}_c$. The reason that the above new code construction is equivalent to the previous description of the repair operations is as follows. Whenever, $\mathbf{v}_c\notin (\mathbf{v}_b\oplus \mathbf{D})$, then the coded packet $Z_c^*$ will not be a linear combination of $Z_b^*$ and the two packets in node $d$. Similarly, whenever $\mathbf{v}_c\notin (\mathbf{v}_b\oplus \mathbf{E})$, then the coded packet $Z_c^*$ will not be a linear combination of $Z_b^*$ and the two packets in node $e$. The choice of $\mathbf{v}_c \in \mathbf{C}\backslash (\mathbf{v}_b\oplus \mathbf{D}))\cup(\mathbf{v}_b\oplus \mathbf{E}))$ thus satisfies both conditions simultaneously.

\par  We argue now that we can always find such a vector $\mathbf{v}_c$. Specifically, we have that
\begin{align}
\rank(\mathbf{C}\cap (\mathbf{v}_b \oplus \mathbf{D}))&=\rank(\mathbf{C})+\rank(\mathbf{v}_b\oplus \mathbf{D})-\nonumber\\
&\quad\quad \rank(\mathbf{C}\oplus(\mathbf{v}_b\oplus \mathbf{D}))\nonumber \\
&\leq  2+3-\rank(\mathbf{C}\oplus(\mathbf{v}_b\oplus \mathbf{D}))\label{eq:rev5}\\
&=2+3-4\label{eq:choose_c}\\
&=1,\nonumber
\end{align}
where \eqref{eq:rev5} follows from $\rank(\mathbf{C})\leq 2$ and $\rank(\mathbf{v}_b\oplus \mathbf{D})\leq 3$. Equation \eqref{eq:choose_c} is due to the following facts. If $\rank(\mathbf{C}\oplus \mathbf{D})=4$, then obviously we have $\rank(\mathbf{C}\oplus\mathbf{v}_b\oplus \mathbf{D})=4$. If $\rank(\mathbf{C}\oplus \mathbf{D})\leq 3$, then by the induction assumptions (A) and (B) we must have $\rank(\mathbf{C}\oplus \mathbf{D})=3$. In this case, we have $\mathbf{S_1}=\mathbf{C}\oplus \mathbf{D}$ when we construct $\mathbf{v}_b$. Therefore, $\rank(\mathbf{C}\oplus \mathbf{v}_b\oplus \mathbf{D})=\rank(\mathbf{C}\oplus \mathbf{D})+1=4$. The above argument shows that $\rank(\mathbf{C}\cap (\mathbf{v}_b\oplus \mathbf{D}))\leq 1$. Symmetrically, we also have $\rank(\mathbf{C}\cap (\mathbf{v}_b\oplus \mathbf{E}))\leq 1$. By a verbatim argument as used in proving $|\mathbf{B}\backslash (\mathbf{S_1}\cup \mathbf{S_2)}|\geq 1$, this implies that $|\mathbf{C}\backslash ( (\mathbf{v}_b\oplus \mathbf{D})\cup (\mathbf{v}_b\oplus \mathbf{E}))|\geq 1$.  Therefore, we can always find such a $\mathbf{v}_c$. 

\par Thus far, we have proven that whenever the induction conditions (A) and (B) hold for time $\tau=\tau_0-1$, we can always carry out the code construction for time $\tau=\tau_0$. We now argue that the induction conditions (A) and (B) also hold after we finish the repair operation in time $\tau=\tau_0$. Since only node $a$ is repaired, we only need to check the node pairs $(a,b)$, $(a,c)$ to make sure they satisfy induction condition (A) and check node pairs $(a,d)$ and $(a,e)$ to make sure they satisfy induction condition (B). 

\par The newcomer $a$ now has packets  $(Z_b^*,Z_c^*)$ and the span of the vectors in $a$ is now $\mathbf{A}=\sspan(\{\mathbf{v}_b,\mathbf{v}_c\})$. By our CA helper selection algorithm, the helper nodes $b$ and $c$ do not form a parent-child relationship. By induction condition (B), $\rank(\mathbf{B}\oplus \mathbf{C})=4$. Thus, any non-zero vector $\mathbf{v}_c\in \mathbf{C}$ is independent of the linear space $\mathbf{B}$. Therefore, $\rank(\mathbf{A}\oplus \mathbf{B})=\rank(\mathbf{v}_c\oplus \mathbf{B})=3$. Symmetrically, $\rank(\mathbf{A}\oplus \mathbf{C})=3$. 

\par To prove that $\rank(\mathbf{A}\oplus \mathbf{D})=4$, we notice that since $\mathbf{v}_b\notin \mathbf{S_1}$ and $\mathbf{S_1}\supseteq \mathbf{D}$, we must have $\mathbf{v}_b\notin \mathbf{D}$. Therefore, $\rank(\mathbf{v}_b\oplus \mathbf{D})=3$. Since $\mathbf{v}_c\notin (\mathbf{v}_b\oplus \mathbf{D})$, we have $\rank(\mathbf{A}\oplus \mathbf{D})=\rank(\mathbf{v}_c\oplus (\mathbf{v}_b\oplus \mathbf{D}))=4$. Symmetrically, we have $\rank(\mathbf{A}\oplus \mathbf{E})=4$. We can thus see that the nodes satisfy the induction conditions (A) and (B) after the repair operations in  $\tau=\tau_0$.

\par We have shown thus far by induction that we can always repair the network/code at any time using the regular repair operations. The above also shows that we can maintain the induction conditions (A) and (B) at any time. We are thus only left with showing that we can construct the whole file from any $k=3$ nodes. Pick any three nodes in the network. By the CA scheme, these three nodes do not form a triangle, i.e., at least one pair of nodes in these three nodes does not form a parent-child relationship. By induction condition (A), we have that the 4 packets on these two nodes are linearly independent and we can use those packets to construct the file. The proof is hence complete.

\section{Proof of Proposition~\ref{prop:mbr}}\label{app:mbr_proof}
To find the MBR point, we need to find the smallest $\beta$ that satisfies
\begin{align}\label{eq:mbr_00}
\min_{\forall \pi_f} \sum_{i=1}^{k}\min \left(\left(d-y_i(\pi_f)\right)^+\beta,\alpha\right)\geq \mathcal{M},
\end{align}
where $\alpha=\infty$ since we are considering the MBR point. One can easily see that the minimzing $\beta$, termed $\beta_\text{MBR}$, equals to $\frac{\mathcal{M}}{\min_{\forall \pi_f} \sum_{i=1}^k(d-y_i(\pi_f))^+}$. Therefore, what remains to be proven is that
\begin{align}
\min_{\forall \pi_f} \sum_{i=1}^k(d-y_i(\pi_f))^+= \sum_{i=1}^k(d-y_i(\pi_f^*))^+ \label{eq:mbr_0}.
\end{align}
where $\pi_f^*$ is the RFIP defined in Section~\ref{subsec:mfhs}. 

\par In our previous work \cite[Proposition~1]{arxiv1}, we have proven the following statements. For any $(n,k,d,r)$ value satisfying $r=0$, we have 
\begin{align}
&\sum_{i=1}^k y_i(\pi_f)\leq \sum_{i=1}^k y_i(\pi_f^*)\label{eq:summary1}\\
&\text{and}~y_{i_1}(\pi^*_f)\geq y_{i_2}(\pi^*_f)~\text{if}~i_1<i_2.\label{eq:summary2} 
\end{align}
Namely, the the RFIP $\pi_f^*$ maximizes the cumulative sum of $y_1(\pi_f^*)$ to $y_k(\pi_f^*)$ and the $y_i(\pi_f^*)$ value is non-decreasing with respect to $i$. 

\par We now argue that \eqref{eq:summary1} and \eqref{eq:summary2} hold for arbitrary $(n,k,d,r)$ value with $r>0$ as well. In the proof of Proposition~\ref{prop:low_b}, the paragraph right before \eqref{eq:marker1}, we have established that the $y_i(\cdot)$ function defined for one scenario $(n,k,d,r)$ is identical to the $y_i(\cdot)$ function defined for another scenario $(n',k',d',r')$ if we have $n=n'$, $d+r=d'+r'$. For any given $(n,k,d,r)$ value, consider another set of parameters $(n',k',d',r')$ such that $n'=n$, $k'=k$, $d'=d+r$, and $r'=0$. Since \eqref{eq:summary1} and \eqref{eq:summary2} hold for any parameter values with $r=0$, they must hold for the case of $(n',k',d',r')$ since by our construction we have $r'=0$. On the other hand, by the arguments in the proof of Proposition~\ref{prop:low_b}, the $y_i(\cdot)$ functions for both $(n,k,d,r)$ and $(n',k',d', r')$ must be identical. Therefore \eqref{eq:summary1} and \eqref{eq:summary2} hold for the arbitrarily given $(n,k,d,r)$ as well. 

\par We now use \eqref{eq:summary1} and \eqref{eq:summary2} to prove \eqref{eq:mbr_0}. Given any $(n,k,d,r)$ value, we construct the corresponding $y_i(\cdot)$ function and the RFIP $\pi_f^*$. Then we define $k_0=\max\{x\in \{1,2,\cdots, k\}: y_x(\pi_f^*)\leq d\}$. Namely, $k_0$ is the largest index $x\leq k$ such that $y_{x}(\pi_f^*)\leq d$. Since by \eqref{eq:summary2} $y_i(\pi_f^*)$ is non-decreasing, we must have $y_i(\pi_f^*)\leq d$ for all $1\leq i\leq k_0$ and $y_i(\pi_f^*)>d$ for all $k_0<i\leq k$. 
 
\par Consider any family permutation $\pi_f$,  we now have
\begin{align}
\sum_{i=1}^{k}(d-y_i(\pi_f))^+&\geq \sum_{i=1}^{k_0}(d-y_i(\pi_f))^+\label {eq:ccw-mbr-3}\\
&\geq \sum_{i=1}^{k_0}(d-y_i(\pi_f))\label {eq:ccw-mbr-4}\\
&\geq \sum_{i=1}^{k_0} (d-y_i(\pi_f^*))\label{eq:mbr_1}\\
&= \sum_{i=1}^k (d-y_i(\pi_f^*))^+\label{eq:mbr_2},
\end{align}
where \eqref{eq:ccw-mbr-3}  follows from that each $(d-y_i(\pi_f))^+$ is non-negative and we sum over $i=1$ to $k_0$ for some $k_0\leq k$; \eqref{eq:ccw-mbr-4} follows from removing the projection $(\cdot)^+$ operator; \eqref{eq:mbr_1} follows \eqref{eq:summary1}; and \eqref{eq:mbr_2} follows by the definition of $k_0$, which ensures $y_i(\pi_f^*)>d$ for all $k_0<i\leq k$. By \eqref{eq:mbr_2}, we get \eqref{eq:mbr_0}. Hence, the proof is complete.
\bibliography{paper}
\bibliographystyle{IEEEtranS}

\end{document}